\documentclass[longauth]{aa}
\usepackage[utf8]{inputenc}

\usepackage{graphicx}
\usepackage{txfonts}
\usepackage[colorlinks, citecolor=blue, breaklinks=true]{hyperref}
\usepackage{longtable}
\usepackage{siunitx}
\usepackage[normalem]{ulem}
\usepackage{xspace}
\usepackage{multirow}
\usepackage{soul}

\def\Msun{$M_\odot$}
\def\Rsun{$R_\odot$}
\def\Lsun{$L_\odot$}
\def\kms{km~s$^{-1}$}
\def\teff{$T_\mathrm{eff}$}
\def\logg{$\log{g}$} 
\def\feh{$\mathrm{\left[Fe/H\right]}$} 

\providecommand{\diam}{\theta_{\rm LD}}

\providecommand{\plato}{\textsc{Plato}}
\providecommand{\gaia}{\textit{Gaia}}
\newcommand{\numax}{\ensuremath{\nu_{\rm max}}\xspace}
\newcommand{\dnu}{\ensuremath{\Delta\nu}\xspace}
\def\simbad{\textsc{Simbad}}
\def\hipparcos{\emph{Hipparcos}}
\def\galah{\textsc{Galah}}
\def\apogee{\textsc{Apogee}}

\def\tess{\textsc{Tess}}
\def\kepler{\emph{Kepler}}

\def\rhostar{\mbox{$\langle\rho\rangle$}}
\def\rhosun{\mbox{$\langle\rho\rangle_{\odot}$}}

\makeatletter 
\let\linenumbers\relax 
\let\runningpagewiselinenumbers\relax 
\makeatother 
\nolinenumbers

\makeatletter 
\renewcommand*\aa@pageof{, page \thepage{} of \pageref*{LastPage}}\makeatother 

\title{\plato\ benchmark stars}
\subtitle{I. The initial benchmark sample}
\titlerunning{\plato\ Benchmark Stars}
\authorrunning{Merle et al.}

\author{
T. Merle\inst{\ref{ulb},\ref{rob}}, 
P. F. L. Maxted\inst{\ref{inst:Keele}},
O. L. Creevey\inst{\ref{oca}},
U. Heiter\inst{\ref{inst:UAO}},
D.B. Palakkatharappil\inst{\ref{inst:IAS}},
K. G. Hełminiak\inst{\ref{inst:NCAC}},
C.~C.~Aguilera~G\'{o}mez\inst{\ref{inst:UCC}},
M. Bazot\inst{\ref{inst:HITS}},
S. Cassisi\inst{\ref{inst:INAF2}, \ref{inst:INFN}},
N. Ebrahimkutty\inst{\ref{oca}},
R. E. Giribaldi\inst{\ref{inst:INAF}},
P. Jofré\inst{\ref{inst:udp}},
F. Kiefer\inst{\ref{inst:LESIA}},
M. N. Lund\inst{\ref{inst:AU}},
N. Miller\inst{\ref{inst:UAO}},
T. Morel\inst{\ref{inst:ULg}},
D. Mourard\inst{\ref{oca}},
N. Nardetto\inst{\ref{oca}},
J.-D.~do~Nascimento,~Jr.\inst{\ref{inst:UFRN}}, 
T. Olander\inst{\ref{inst:UAO}},
G. N. Pawar\inst{\ref{inst:NCAC}},
A.~Rebassa-Mansergas\inst{\ref{inst:Catalunya1},\ref{inst:Catalunya2}},
R.~Raddi\inst{\ref{inst:Catalunya1}},
B.~Rojas-Ayala\inst{\ref{inst:IAI}},
J.~Southworth\inst{\ref{inst:Keele}}
}
\institute{
BLU-ULB, Institut d'Astronomie et d'Astrophysique, Universit\'e Libre de Bruxelles, CP 226, Bvd du Triomphe, Belgium, \\ \email{thibault.merle@ulb.be}, \url{www.ulb.be}  \label{ulb} \and
Royal Observatory of Belgium, Avenue Circulaire 3, 1180 Brussels, Belgium\label{rob} \and
Astrophysics group, Keele University, Staffordshire, ST5 5BG, UK\label{inst:Keele} \and
Universit\'e C\^{o}te d’Azur, Observatoire de la C\^{o}te d’Azur, CNRS, Laboratoire Lagrange, France\label{oca}  \and
Observational Astrophysics, Department of Physics and Astronomy, Uppsala University, Box 524, SE-751 20 Uppsala, Sweden\label{inst:UAO} 
\and
Institut d'Astrophysique Spatiale (IAS),  CNRS, Université Paris-Saclay, Orsay,91400, France \label{inst:IAS}
\and Stellar Astrophysics Centre, Department of Physics and Astronomy, Aarhus University, Ny Munkegade 120, Denmark\label{inst:AU} \and
Nicolaus Copernicus Astronomical Center, Polish Academy of Sciences, ul. Rabia\'{n}ska 8, 87-100 Toru\'{n}, Poland\label{inst:NCAC} \and 
Instituto de Alta Investigaci\'on, Universidad de Tarapac\'a, Casilla 7D, Arica, Chile\label{inst:IAI} \and
INAF – Osservatorio Astrofisico di Arcetri, Largo E. Fermi 5, 50125
Firenze, Italy\label{inst:INAF}
\and STAR Institute, Université de Liège, Quartier Agora, Allée du 6 Août 19c, Bât. B5C, 4000 Liège, Belgium\label{inst:ULg}
\and LESIA, Observatoire de Paris, Université PSL, CNRS, Sorbonne Université, 5 place Jules Janssen, 92195 Meudon, France \label{inst:LESIA}
\and Heidelberger Institut für Theoretische Studien, Schloss-Wolfsbrunnenweg 35, 69118 Heidelberg, Germany \label{inst:HITS}
\and
Instituto de Estudios Astrofísicos, Universidad Diego Portales, Av. Ejército Libertador 441, Santiago de Chile\label{inst:udp}
\and
Departament de F\'{\i}sica, Universitat Polit\`{e}cnica de Catalunya, c/ Esteve Terrades 5, 08860 Castelldefels, Spain\label{inst:Catalunya1}
\and
Institut d'Estudis Espacials de Catalunya, c/ Esteve Terradas, 1, Edifici RDIT, Campus PMT-UPC, 08860 Castelldefels, Spain\label{inst:Catalunya2}
\and
INAF – Astronomical Observatory of Abruzzo, Via M. Maggini, sn, 64100 Teramo, Italy\label{inst:INAF2}
\and
INFN, Sezione di Pisa, Largo Pontecorvo 3, 56127 Pisa, Italy\label{inst:INFN}
\and
Instituto de Astrofísica, Pontificia Universidad Católica de Chile, Av. Vicuña Mackenna 4860, 782-0436 Macul, Santiago, Chile \label{inst:UCC}
\and
Universidade Federal do Rio Grande do Norte (UFRN), Departamento de F\'isica, 59078-970, Natal, RN, Brazil \label{inst:UFRN}
}

\date{Received; accepted}

\abstract
{The ESA \plato\ mission aims to detect and characterise Earth-size exoplanets orbiting bright, late-type stars through high-precision photometry and asteroseismology. Precise and accurate stellar parameters of host stars are essential for reliable exoplanet characterization and benchmarking of stellar models.}
{This work defines a comprehensive sample of \plato\ benchmark stars with well-determined fundamental and atmospheric properties that satisfy stringent precision requirements to support mission pre-operation, operation, and post-operation phases.}
{We prepared a catalogue of benchmark stars by compiling data from diverse sources, including eclipsing and spectroscopic binaries, interferometric angular diameters, asteroseismic measurements, and large spectroscopic and photometric surveys. Fundamental and atmospheric stellar properties (mass, radius, luminosity, age, effective temperature, gravity, density and limb darkening) were assigned benchmark levels reflecting their degree of model dependence. We also provide ancillary parameters like parallax, metallicity, bolometric flux, angular diameter, extinction, rotational period; and X-ray fluxes (the latter uniquely for M dwarfs). Parallax data were homogenized and when \gaia~DR3 parallaxes are used, we systematically took into account the parallax bias. Multiple independent measurements were reconciled to produce a coherent and robust set of stellar properties adhering to the \plato\ mission’s precision and accuracy requirements.}
{The \plato\ benchmark catalogue contains 1090 FGKM main-sequence and subgiant stars, divided into three quality tiers based on model-independent (B1) parameters: \emph{gold} (15 stars with B1 mass, radius, and luminosity), \emph{silver} (195 with B1 mass and radius), and \emph{bronze} (416 with at least one B1 property). The remaining stars lack B1 parameters but feature well-characterised model-dependent (B2) or empirical (B3) properties, expanding the parameter space for model validation. The stars are well-distributed in sky position and brightness, making them optimally suited for testing and calibration of \plato\ pipelines and data. Approximately 40\% of the stars in the catalogue lie within the \plato\ fields of view. The catalogue provides homogenized stellar parameters with rigorous uncertainty estimates, enabling validation of \plato\ data processing pipelines, empirical testing of stellar evolutionary and asteroseismic models, and cross-survey calibration.}
{This \plato\ benchmark star sample (PBS, v1.0) constitutes the largest catalogue of benchmark stars to date and a fundamental resource anchoring \plato’s scientific goals in exoplanet and stellar astrophysics. By providing rigorously vetted benchmark stars, the mission ensures accurate planetary parameter determinations and advances our understanding of stellar structure and evolution. The catalogue is available online at \url{https://astro.ulb.ac.be/pbs} and will be updated regularly, offering long-term value to the \plato\ project and the broader astronomical community.}


\begin{document}

\maketitle

\section{Introduction}
The ESA \plato\ mission \citep[PLAnetary Transits and Oscillation of stars,][]{rauer2014, rauer2025} is scheduled to be launched in January 2027. It is designed to detect and characterise large numbers of planets orbiting nearby stars, including Earth-size planets orbiting in the habitable zone of Sun-like stars.
Mean densities and ages will be determined for the terrestrial planets. The exoplanet densities will be computed using radii obtained from the \plato\ transit light-curves and masses from ground-based radial velocity follow-up observations. Masses, radii and ages of FGK-type host stars will be characterised by asteroseismic analyses of the \plato\ photometric light-curves \citep[\emph{e.g.}][]{goupil2024}. The properties of all types of host stars, including M dwarfs, will be determined from ground-based preparatory, archival, and follow-up observations. \plato's unique capability of asteroseismic analysis and systematic observation of bright solar-type and M-dwarf stars will enable unprecedented precision in stellar parameters for thousands of stars hosting terrestrial exoplanets, fundamentally advancing our knowledge of planetary system architectures and stellar evolution.

This justifies defining a sizeable \plato\ sample of benchmark stars (hereafter PBS) in a similar manner to that used for the \gaia\ benchmark stars \citep[GBS v3,][]{soubiran2024}, building upon \citet{heiter2015AA...582A..49H}.
Such sample will serve as essential references for a variety of scientific cases and for the validation and calibration of analysis methods during all phases of the \plato\ mission, including pre-operation (PO), operation (OPE), and post-operation. 
The scientific motivation for assembling such a sample is anchored in the stringent requirements set by \plato\ for the precise and accurate determination of stellar masses, radii, luminosities, ages, effective temperatures, surface gravities, and metallicities across a range of FGKM main-sequence and sub-giant stars. 
These high-fidelity benchmarks provide the foundation for:

\begin{itemize}
    \item \textit{Validation of automated pipelines:} robust benchmark stars allow the testing and cross-calibration of pipelines for atmospheric parameter determination, asteroseismic modelling, and exoplanet transit analysis, ensuring that the extracted stellar parameters are reliable and consistent throughout the mission.
    \item \textit{Cross-survey calibration:} by anchoring \plato's results to a common set of well-characterised stars, it becomes possible to cross-calibrate with legacy surveys and ongoing ground-based spectroscopic and space-based photometric surveys, thus facilitating the homogenisation of stellar parameters on a community-wide scale.
    \item \textit{Precise exoplanet characterisation:} accurate benchmark stars are crucial for determining host star properties, which directly influence the inferred radii, densities, and ages of discovered exoplanets \citep[e.g.][]{stassun2017, stassun2018}; fundamental steps for addressing key questions on planetary composition, formation, and habitability.
    \item \textit{Stellar physics and evolution:} the availability of stars with independently determined parameters serves as a critical testbed for stellar evolutionary models, empirical relations, and the exploration of physical processes across different evolutionary stages. Ultimately, such benchmark stars are essential for refining the next generation of stellar models required by the \plato\ mission.
\end{itemize}

While existing state-of-the-art compilations, such as the GBS series, provide invaluable reference standards, they are limited in size and do not sufficiently cover the specific parameter space, brightness ranges, or M-dwarf demographics required to validate the automated pipelines of the \plato\ mission. This highlights the critical need for a dedicated, homogeneous benchmark sample tailored specifically to \plato's core scientific specifications.

The WP125500 ``Benchmark stars'' work package was established in May 2018 to organise the compilation of fundamental stellar properties, guided by requirements from the leaders of the stellar science work packages (WP121--128). Over several years, members of WP125500 gathered extensive literature measurements. Distilling this heterogeneous compilation into a single, curated, high-quality database with one definitive entry per star represented a major validation and software development effort.

Furthermore, the PBS sample is an essential component of the mission's science calibration and validation plan. The part of the PBS located in the \plato\ field-of-view (FOV) selected for the first two years of the mission (see Sect.~\ref{sec:cat-gen}) has already been included in the science calibration and validation subset of the \plato\ input catalogue (scvPIC, \citealt{zwintz2026}).
A representative example includes the use of the PBS to refine asteroseismic scaling relations, improve the modelling of limb-darkening for transit fitting, and assess systematic errors in stellar radii that propagate to planetary studies.

This paper presents the work compiling the \plato\ benchmark catalogue, and it is structured as follows: Sect.~\ref{sec:properties} introduces the definitions of the stellar properties to be assessed, as well as the ancillary parameters considered in this study. The compilation of input data sets is described in Sect.~\ref{sec:data}. In Sect.~\ref{sec:methods}, we detail the methodology adopted to construct the \plato\ benchmark star catalogue. Finally, Sect.~\ref{sec:cat} presents the resulting catalogue and demonstrates its usage for various scientific applications.

\section{Stellar properties and parameters}
\label{sec:properties}

\subsection{Definition of benchmark levels}
A \emph{benchmark star} is a star for which reliable measurements of the star’s properties are known. 
This implies that any possible sources of systematic errors in the measurements are well understood and, if possible, removed; and that the precision of the property meets the specifications and requirements from the \plato\ stellar science work packages. 

With the goal of performing a quantitative assessment of the reliability of properties, we define three \emph{benchmark levels} for each stellar property: B1, B2 and B3, as defined below. 
We note that these benchmark levels apply to the properties of the star separately, not to the star itself, e.g.  a star may have a measurement of its mass at benchmark level B1, a radius measurement at benchmark level B2 and an age estimate at benchmark level B3.
Note the distinction here between values with a B1 benchmark level that are ``measurements'' whereas values with a B2 or a B3 benchmark level are ``estimates''.

We note that this approach represents a different strategy compared to the GBS \citep{soubiran2024}, which strictly required B1-level parameters across the board to validate automated pipelines. The PBS sample in turn tracks eight parameters independently per star, allowing this flexibility to accommodate a large, diverse sample while maintaining rigorous pipeline validation standards.

\paragraph{Benchmark level B1 -- Model-independent properties.} A direct, accurate and precise measurement of a star's fundamental property with negligible model dependence or coming from first principles. It can be summarised as: 
\emph{``the set of the model assumptions is well defined''}. 
For instance, the mass of a star coming from the orbital dynamics in a visual and double-lined spectroscopic binary is B1.

\paragraph{Benchmark level B2 -- Model-dependent properties.} A property inferred using stellar models or other theoretical inputs where the level of systematic errors due to the model dependence is well understood.  It can be summarised as: \emph{``the set of the model assumptions is loosely defined''}. For instance, the mass of a star is B2 if (i) the star belongs to a cluster and its mass is fitted on the cluster isochrone either using magnitude and colour or luminosity and effective temperature; or (ii) if the stellar mass is inferred from an asteroseismic modelling analysis. 

\paragraph{Benchmark level B3 -- properties from empirical relations.} Assigned when the stellar property is obtained from a phenomenological relation using a mathematical functional form to relate various parameters. It can be summarised as: \emph{``the model is empirical''}. For instance, the mass of a star is B3 if the mass is coming from the mass-luminosity, mass-radius or other similar kind of empirical relations, which are precisely defined for main sequence stars, while discrepancies appear when the stars evolve out of the main sequence. 

\subsection{Stellar properties}
In this work, we assess eight key stellar properties for the \plato\ benchmark stars: (i) the stellar mass, $M$; (ii) the stellar radius, $R$; (iii) the stellar luminosity, $L$; (iv) the stellar age, $\tau$; (v) the effective temperature, \teff; (vi) the surface gravity expressed as the decimal logarithm, \logg; (vii) the mean stellar density expressed as the decimal logarithm, $\log\rhostar$; and (viii) the limb-darkening characterised by the parameters $h^{\prime}_1$ and $h^{\prime}_2$ defined below. Stellar properties are either sourced from the literature or calculated via the equations in Appendix~\ref{app:params}. The values of $M$, $R$, $L$ and $\rhostar$ are given in solar units using the nominal solar values given in Table~\ref{tab:ref}. Several ancillary parameters are included, such as the overall metallicity [Fe/H], but these are not assigned benchmark levels.
 
\subsubsection{The mass}
Mass is the primary property that governs the evolution and ultimate fate of a star. In a recent comprehensive review on the determination of stellar masses, \citet{serenelli2021} present the mass ladder in their Fig.~16 that gives the accuracy/precision as a function of methods used and mass ranges. 

We defined as B1 the mass from a star belonging to a binary system which is observed: 
(i) as a visual and spectroscopic binary (e.g. \citealt{pourbaix2016}) or (ii) as a detached eclipsing and double-lined spectroscopic binary (e.g. \citealt{maxted2020}). In case (i), the sum and ratio of the masses can be obtained from Newtonian dynamics, with the only assumption that the masses are point sources. In case (ii), the inclination of the orbit on the sky obtained from the eclipses allows one to obtain the individual stellar masses from each mass function. While standard binary orbit solutions often assume an isolated system with purely Keplerian motion, tertiary gravitational perturbations or light-travel time effects can be explicitly incorporated when a third companion is present. A clear example is the eclipsing binary AI~Phe, where three-body orbital effects were accounted for in the spectroscopic analysis by \citet{2019AA...632A..31G}.

A stellar mass is considered as B2 when stellar models are used: (i) from interpolation in stellar evolutionary tracks (e.g. \citealt{lebreton2020}),  which is applicable to both cluster members and isolated field stars; (ii) from asteroseismic frequency fitting (e.g. \citealt{metcalfe2023} or from LEGACY sample \citealt{lund2017, silva_aguirre2017}); or from similar methods based on stellar models. 
While both field and cluster stars modelled with evolutionary tracks are classified as B2 (case i), isolated main-sequence stars suffer from highly degenerate and uncertain mass estimates that can only be resolved through the tighter physical constraints of cluster coevality or asteroseismology.

A mass is assigned B3 in the following other cases: (i) when mass is derived from a mass-luminosity, mass-radius, mass-temperature or any other relation (e.g. \citealt{eker2018,moya2018}, or \citealt{benedict2016,mann2019} specifically for M dwarf stars); (ii) from a single-lined spectroscopic binary and other constraints (e.g. \citealt{pourbaix2003, doyle2011}); (iii) from an asteroseismic scaling relation that depends on the frequency of maximum power ($\nu_{\max}$), the large frequency separation ($\Delta \nu$) and effective temperature (e.g. \citealt{hekker2020}).

\subsubsection{The radius}
The term stellar radius generally refers to the Rosseland radius, defined as the radius at which the Rosseland mean optical depth is $2/3$. This physical definition is standard in stellar atmosphere models, but differs in detail from the radius inferred from direct angular diameter measurements from interferometry \citep{brown1968}, or that measured in eclipsing binaries.
However, for main-sequence and subgiant stars, the difference is  very small ($\la 0.1\%$), and is thus usually neglected in practice. 
In special cases requiring high precision, this distinction may become relevant (see, e.g., \citealt{miller2020}). 
For M dwarfs this has not been established to the same precision, and even in the most favourable case (NIR observations, where molecular-opacity effects are minimized) the difference reaches $\sim 0.1-0.2\%$ \citep{2014MNRAS.438.2413V, Rabus2019, 2019A&A...625A..68S}.
We defined as B1 a stellar radius coming from (i) an interferometric angular diameter and an astrometric parallax (e.g. \citealt{kervella2017}) or (ii) from a detached eclipsing and double-lined spectroscopic binary (e.g. \citealt{maxted2020}). 
Similarly to B2 masses, B2 radii are the ones relying on interpolation in stellar evolutionary tracks or asteroseismic frequency fitting. 
Similarly to B3 masses, B3 radii are from similar empirical relations, single-lined spectroscopic binary and other constraints, or from asteroseismic scaling relations. 

\subsubsection{The luminosity}
The stellar luminosity is a property difficult to accurately derive as it requires as many flux measurements as possible in different photometric bands and relies on flux calibrations \citep{bless1997, bohlin2014,soubiran2024}. The luminosity is assigned B1 only when coming from spectrophotometric fluxes and accurate parallaxes, assuming that the flux measurements cover well the whole wavelength range which contains significant contribution to the spectral energy distribution of main-sequence FGKM dwarfs and subgiants (from UV to IR), e.g. \cite{soubiran2024}. It is assigned B2 when the stellar luminosity is derived from: (i) classical photometry and bolometric corrections relying on effective temperature, gravity and metallicity parameters (e.g. \citealt{creevey2012a}); (ii) interpolation in stellar evolutionary tracks (e.g. \citealt{borde2020}); or (iii) asteroseismic frequency fitting and photometry (e.g., \citealt{creevey2012b}). The luminosity is B3 if the surface brightness-colour relations (SBCRs) for the angular diameter is used with parallax and effective temperature. Alternatively, combining empirical bolometric fluxes with high-precision parallaxes offers a powerful, largely model-independent pathway to derive luminosities and radii that can be validated directly against eclipsing binary and interferometric benchmarks \citep{stassun2016, stassun2017}.

\subsubsection{The age}
The age is probably the most difficult stellar property to accurately derive as it is the least direct accessible from the observations. A very complete review on age-dating techniques is given by \citet{soderblom2010}, and on accuracy of ages from stellar models by \citet{2014EAS....65...99L, 2014EAS....65..177L}. See also \citet{2018IAUS..334...11C} for the oldest systems. We find that only the Sun can be considered to have a B1 benchmark age from meteorite age dating and helioseismology \citep{connelly2017,bonanno2015}. All the methods otherwise heavily depend on model assumptions. A B2 age can be derived using (i) cosmochronology in very metal-poor stars using the abundance ratio U/Th (e.g; \citealt{cayrel2001, shah2023}), (ii) isochrone fitting in stellar clusters (e.g. \citealt{edvardsson1993}), (iii) asteroseismic frequency fitting (e.g. \citealt{metcalfe2023}), (iv) white dwarf cooling in wide binaries based on an initial-to-final mass relation (e.g. \citealt{rebassa2021, 2026A&A...709A.250R}), or (v) the Li depletion boundary (e.g. \citealt{jeffries2013}). An age is B3 if based on (i) an asteroseismic scaling relation (e.g. \citealt{creevey2017}), (ii) the gyrochronology method (e.g. \citealt{barnes2007}), (iii) the activity from chromospheric or coronal emission lines (e.g. \citealt{mamajek2008}), (iv) from abundance ratio-age relations (e.g. \citealt{nissen2020}), or (v) membership of a stellar population (e.g. \citealt{2023ApJ...954L..50E}).

\subsubsection{The effective temperature}
The effective temperature, \teff, for a star with Rosseland radius $R$ and luminosity $L$ is defined by the equation  
\begin{equation}
L=4\pi R^2 \sigma_{\rm SB} T_{\mathrm{eff}}^4,
\end{equation}
where $\sigma_{\rm SB}$ is the Stefan-Boltzmann constant.\footnote {Note that this definition says nothing about the spectrum of the light emitted by the star, i.e. \teff\ is the effective temperature of a star whether or not it emits light as a black body.} 
Dividing both sides of the equation by $d^2$, where $d$ is the distance to the star, leads to the alternative and equivalent definition
\begin{equation}
T_{\mathrm{eff}} \equiv \left(f_{\rm bol}/\sigma_{\rm SB}\right)^{0.25}\theta^{-0.5},
\end{equation}
where $f_{\rm bol}$ is the bolometric flux at the top of the Earth's atmosphere corrected for interstellar extinction and $\theta = 2R/d$ is the angular diameter. The effective temperature is the most important property governing the formation of stellar spectra. Various techniques to estimate \teff\ are summarised in the Supplemental Text 1 of \citet{jofré2019}. 

The values of \teff\ in this compilation are labelled B1 if derived from any of the following three methods: (i) interferometric angular diameter and spectrophotometric fluxes (e.g. \citealt{boyajian2012, soubiran2024}); (ii) eclipsing binary radius with orbital parallax and spectrophotometric fluxes (e.g. \citealt{2022MNRAS.513.6042M}); (iii) differential spectroscopic analysis with respect to the Sun only for solar twins with \teff\ within 50~K of the Sun, and \logg, and [Fe/H] within 0.05~dex of the Sun (e.g. \citealt{melendez2012}). \teff\ is assigned a benchmark level of B2 when estimated from: (i) fitting H$\alpha$ wing profiles using synthetic spectra from 3D and non-LTE stellar atmosphere models (e.g. \citealt{giribaldi2021}); (ii) using the infrared flux method (IRFM) based on an absolute calibration of Vega (\citealt{casagrande2006}); (iii) broad-band photometric colours through \teff-colour relations (e.g. \citealt{vandenberg2003}); (iv) differential analysis with stellar analogues having B1 \teff; (v) spectroscopic excitation equilibrium of neutral iron lines \citep{jofré2019}; or (vi) any other fitting method with spectral syntheses \citep[e.g.][]{2014AA...567A..64H}. Finally, \teff\ is B3 if based on empirical relations, e.g. \teff\ as a function of  mass for main-sequence stars \citealt{eker2018, moya2018}).

Until recently, interferometric measurements have been the main source of direct angular diameter measurements for dwarf stars. A Sun-like star at a distance of 50\,pc has an angular diameter $\theta \la 0.2$\,mas, which corresponds to the resolution limits of modern visible-light long-baseline optical interferometers \citep{2005ApJ...628..453T,2013JAI.....240002A,2010SPIE.7734E..04H,mourard2026}. Thanks to recent instrumental advancements, significant progress has been made in understanding, controlling, and mitigating measurement biases and systematic errors. Nevertheless, repeated measurements of $\theta$ from different studies of the same star have historically shown differences of $\ga 5$\,per~cent \citep{soubiran2024}, even when using the same underlying data \citep{bond2020}, see discussion in Sect.~\ref{sec:caveats} for more details.

The advent of the \gaia\ mission opens up the possibility to tackle this problem by using angular diameter measurements for stars in eclipsing binary systems computed from $R/d$ where the radius $R$ is measured from the analysis of the light curve and the spectroscopic orbit, and the distance $d$ is inferred from the parallax. With space-based photometry and radial velocities measured using stabilised \'echelle spectrographs, it is now possible to measure $R$ with an accuracy $\la 0.5$\,per~cent for stars in bright eclipsing binary systems with narrow total eclipses \citep{maxted2020,2021A+A...649A.109G}. This makes it possible to accurately measure \teff\ with a precision $\la 50$\,K for stars in detached eclipsing binary systems given some method to estimate $f_{\rm bol}$ for the two stars independently from their observed combined fluxes and directly measured flux ratios \citep{miller2020,2025MNRAS.544.4611M}. The number of stars in eclipsing binaries that have direct \teff\ measurements using this method is currently quite small, but is growing rapidly. Thus, we can hope to soon have a large enough sample to establish an accurate zero-point for the effective temperature estimates for solar-type stars using spectroscopic and photometric methods. 

\subsubsection{The surface gravity}
The strength and width of absorption lines in the stellar spectra are affected by the local pressure in the line-forming region. For a star in hydrostatic equilibrium, this pressure is directly related to the surface gravity, $g$. By convention, the logarithm of surface gravity is denoted \logg, with $g$ in {\it cgs} units ($\mathrm{cm\,s^{-2}}$), so \logg $\approx 4$ for dwarf stars. Thus, an accurate estimate for \logg\ is needed to obtain accurate abundance estimates from the analysis of stellar spectra. It is possible to estimate \logg\ from the analysis of the spectrum itself, but the correlation between \logg\ and abundance together with the approximations used in stellar model atmospheres, instrumental effects, etc. can lead to systematic errors $\ga 0.2$~dex in estimates of \logg\ from spectroscopy \citep{2018A&A...620A..76A}. These issues motivate independent \logg\ determinations. For example, the APOKASC collaboration \citep{2018ApJS..239...32P} leveraged \kepler\ asteroseismic gravities to calibrate systematic spectroscopic offsets in \logg\ estimates from APOGEE.

The various techniques to determine \logg\ are summarized in the Supplemental Text 1 of \citet{jofré2019} (see also Sect.~2 of \citealt{creevey2013}). Benchmark level B1 is attributed to \logg\ if determined from a B1 mass and a B1 radius. It is B2 if it comes from: (i) mass and radius where one is B2 and the other is B1 or B2 (including values derived from grid-based asteroseismic modeling of individual frequencies); (ii) a primary star in an eclipsing binary with an SB1 orbit, determined via its mean stellar density inferred from light curves combined with a primary mass estimate (e.g. \citealt{maxted2023}); (iii) fitting the wings of strong lines if the metallicity is known \citep[Supplemental Text~1 of ][]{jofré2019}; or (iv) spectroscopic ionisation equilibrium from neutral and ionised iron lines. It is assigned B3 if it is computed directly from the asteroseismic $\nu_{\text{max}}$ scaling relation (Eq.~\ref{eq:numax}) without detailed grid modeling, or if it is derived from a mass and radius where at least one is B3.

\subsubsection{The mean stellar density}
\label{sec:rhostar}
Throughout this work, we report the mean stellar density as its decimal logarithm, $\log\langle\rho\rangle$, in solar mean density units ($\rho_\odot = 1410~\mathrm{kg\,m^{-3}}$).
For any star of mass $M$ and radius $R$ in a detached binary system with a companion of mass $q M$, we can use Kepler's 3$^{\rm rd}$ law to show that the mean stellar density is given by
\begin{equation}
\label{eqn:rhostar}
  \langle\rho\rangle = \frac{3M}{4\pi R^3} =    \frac{3\pi}{GP^2(1+q)}    \left(\frac{a}{R}\right)^3,
  \end{equation}
where $P$ is the orbital period and $a$ is the semi-major axis of the orbit. This quantity $a/R$ is closely related to the duration of the transit in binary stars with total eclipses, and in transiting exoplanet systems. This means that the precision of planet radius estimates from the analysis of their transits can be significantly improved by including an independent estimate of $\langle\rho\rangle$ in cases where the light curve is too noisy to yield an accurate value of the transit impact parameter. This is of particular relevance to the analysis of \plato\ light curves for Earth-size planets orbiting in the habitable zone of Sun-like stars. It also means that we can make accurate estimates of \rhostar\ from high signal-to-noise light curves of transiting exoplanets and eclipsing binary systems with very low mass companions (EBLM systems) without the need for accurate, independent estimates for $M$ and $R$, but this method does require a precise estimate for the eccentricity of the orbit \citep{2025MNRAS.539.3741M}. Alternatively, Eq.~(\ref{eqn:rhostar}) illustrates how the star-planet connection can be leveraged: an independent, precise determination of the stellar density $\langle \rho \rangle$ (e.g. from asteroseismology; \citealt{van_eylen2014,lund2019}), combined with the transit-derived scaled semi-major axis $a/R$, provides a powerful geometric constraint on the system's orbital eccentricity. The dependence of \rhostar\ on $(1+q)$ can be ignored in systems where $q\rightarrow 0$, so measurements of \rhostar\ from the light curves of transiting exoplanets with accurate and independent eccentricity measurements are assigned a benchmark level of B1. For typical EBLM systems with $q\approx 0.2$, an error of 10\% in the estimate of  $M$ implies an error $\lesssim 1$\% in $\rhostar$. Although this correction is small, it may not be negligible in cases where very accurate values of \rhostar\ are needed, so measurements of \rhostar\ from the light curves of EBLM systems are assigned a benchmark level of B2. In cases where independent estimates of the stellar mass are preferred to the values given in the database for the EBLM systems, a small correction can be applied to the values of \rhostar\ and \logg. The values of $d\log(\rhostar)/dM$ needed for this correction can be found in \citet{2025MNRAS.539.3741M}. Consequently, while no direct observational measurement method is formally classified at the B3 level for mean stellar density, a B3 benchmark level is assigned to 51 targets in our compiled catalogue (predominantly within our M-dwarf sample) where the calculated density inherits the benchmark level of an empirically-derived B3 mass.

\subsubsection{The limb darkening}
The reduction in specific intensity from the centre to the limb of the stellar disc is known as limb darkening.
The specific intensity emitted from a stellar photosphere is typically denoted $I_{\lambda}(\mu)$, where
$\mu$ 
is the cosine of the angle between the line of sight and the surface normal vector and $\lambda$ is the wavelength or bandpass of the observations under consideration. 
For a spherical star, $\mu = \sqrt{1-r^2}$, where $r$ is the radial coordinate from $r=0$ at the centre of the stellar disc to $r=1$ at the limb. 
Limb-darkening can be seen directly in high-quality light curves of stars that show transits from an exoplanet or other dark body (e.g. a brown dwarf or very low mass star) as the curvature between the second and third contact points as the companion scans across the non-uniform intensity distribution of the stellar disc. 

\citet{2023MNRAS.519.3723M} defined the following parameters that can be used to directly compare limb-darkening profiles measured from high-quality light curves of transiting exoplanets to the specific intensity profile computed from stellar model atmospheres:
\begin{equation}
\begin{array}{ll}
h^{\prime}_1 = & I_{\lambda}(2/3);\\
\noalign{\smallskip}
h^{\prime}_2 = & h^{\prime}_1 - I_{\lambda}(1/3).
\end{array}
\end{equation} 
Using these parameters that are directly related to the specific intensity profile avoids the ambiguities that arise from using coefficients of limb-darkening laws measured from light curves as benchmark quantities. Simulations described in \citet{2023MNRAS.519.3723M} show that the values of ($h^{\prime}_1,h^{\prime}_2$) obtained from the least-squares fit of a 4-parameter limb-darkening law are accurate, i.e. close to the actual values for the assumed limb-darkening profile used for the simulations, and that these values are not strongly correlated with each other if the transit impact parameter is not too large ($b \lesssim 0.6$). It was also found that the values of ($h^{\prime}_1,h^{\prime}_2$) obtained were almost identical if a different limb-darkening law was used for the least-squares fits. Based on these results, we consider the values of ($h^{\prime}_1,h^{\prime}_2$) measured from high signal-to-noise transit light curves for systems with $b \lesssim 0.6$  using a fit of a flexible limb-darkening law to be benchmark level B1. The values in the benchmark database are taken from \citet{2024MNRAS.534.3893V} and \citet{2025MNRAS.537.3943F}. \citeauthor{2024MNRAS.534.3893V} analysed the same sample as  \citet{2023MNRAS.519.3723M} using a regularisation technique to improve the precision of the results. 

\subsection{Ancillary parameters}
Ancillary parameters are also included in the compilation. These were used to derive or update values of the stellar properties given the in database, or to provide additional information about the benchmark stars. These ancillary parameters are the parallax $\varpi$, the limb-darkened angular diameter $\diam$\footnote{The angular diameter values used to calculate stellar properties are those quoted values taken from the cited works, for which limb-darkened models were used to interpret the observations obtained from Earth.}, the bolometric flux $f_\mathrm{bol}$, the colour excess $E(B-V)$, the rotation period $P_\mathrm{rot}$, the iron abundance relative to the Sun [Fe/H] (the metallicity), and the abundance of $\alpha$-elements relative to iron and to the Sun [$\alpha$/Fe]. In addition, for a sub-sample of M-dwarf benchmark stars a table with X-ray- to bolometric flux ratios $f_X/f_\mathrm{bol}$ is provided (see Sect.~\ref{sec:methods-Md}). Contrary to stellar properties no quality assessment was performed for these ancillary parameters. However, for parallaxes we perform a homogenisation step (see Sect.~\ref{sec:parallaxes}).

\subsection{\plato\ benchmark requirements}
\label{sec:requirements}
The various groups in charge of the analysis of \plato\ data agreed to test and validate their software on 
a benchmark sample comprising a few hundred FGKM main-sequence and sub-giant stars. The requirements on the range and precision of the stellar properties and parameters for this sample are summarised here:
\begin{itemize}
 \item masses in the range [0.08, 2.0]~\Msun\ with $\Delta M/M\le5$\%
 \item radii with $\Delta R/R\le 5$\%
 \item luminosities with $\Delta L/L \le 5$\%
 \item age with $\Delta \tau/\tau\le 10$\%
 \item \teff\ in the range [2800, 6700]~K with $\Delta T_\mathrm{eff}\le 100$~K
 \item \logg\ $\ge$ 3.5\footnote{We follow the usual convention of denoting the logarithm of the surface gravity as \logg\  with $g$ in units of $\mathrm{cm\,s^{-2}}$.} dex with $\Delta \log{g} \le 0.1$~dex 
 \item metallicity in [$-1.5$, +0.5]~dex with $\Delta \mathrm{[Fe/H]}\le0.1$~dex.
\end{itemize}

For mass and \teff, we combine the requirements from the core \plato\ stellar samples (P1, P2, and P5), which apply to F5- to K7-type stars, and include the requirements on late-type M dwarfs from the P4 sample as defined in the \plato\ Input Catalogue (PIC).\footnote{The PIC2.2 \citep{2026arXiv260403369M} categorises targets into several scientific samples: P1 contains at least 15\,000 dwarf and subgiant F5--K7 stars with $V \le 11$~mag; P2 is a bright subset of P1 containing 1\,000 targets with $V \le 8.5$~mag; P5 is the main statistical sample containing at least 245\,000 F5--K7 dwarfs and subgiants with $V \le 13$~mag; and P4 contains at least 5\,000 M-dwarfs with $V \le 16$~mag in the solar vicinity \citep[see][for details]{rauer2025}.} We note that these requirements do not need to be fulfilled simultaneously. Furthermore, for some of the parameters requirements for higher precision than stated in the list were provided by some of the analysis groups. 
 We thus reduce the list above to one that focuses on the most stringent constraints:
\begin{itemize}
\item masses with $\Delta M/M\le1$\%
 \item radii with $\Delta R/R\le1$\%
 \item \teff\ with $\Delta T_\mathrm{eff}\le 50$~K
 \item \logg\ with $\Delta \log{g} \le 0.05$~dex
\end{itemize}

Requirements also consider testing the accuracy of stellar models, which includes the reliability of stellar structure and evolution calculations and the impact of certain choices in the model physics.  These have a particularly important impact on the determination of stellar ages. 
To test the physical accuracy of stellar models, we must examine stars beyond the core subgiant branch. We therefore include a few red giants with model-independent radii from interferometry and masses from binaries. Extending the PBS sample beyond \plato's core demographics (FGKM dwarfs and subgiants) to include giants and metal-poor stars, specifically using the metal-poor dwarf and giant catalogues of \citet{giribaldi2021, giribaldi2023}, is a deliberate strategy. This expanded parameter space provides a larger, robust pool of benchmark-quality properties, offering critical opportunities to validate the \plato\ data processing pipelines under extreme stellar regimes.

\section{Data compilation}
\label{sec:data}
Following an extensive literature compilation by the work package members, our benchmark targets were selected from a combination of diverse published works, which must be combined and complemented to fully satisfy \plato's calibration requirements. As a consequence, the choice of the selected targets is necessarily biased, mainly driven by the \plato\ benchmark requirements. The main sources of data are summarized in Table~\ref{tab:data_compilation_summary}.

\subsection{Binary and multiple systems} 
\label{sec:binary}
To assemble the anchor stars for mass and radius, we rely heavily on binary and multiple stellar systems. 
We made a selection of stars in eclipsing binary systems from version 2025/05/27 of DEBCat\footnote{\url{https://www.astro.keele.ac.uk/jkt/debcat}} \citep[Detached Eclipsing Binary Catalogue,][]{southworth2015} guided by the need to validate the grid of stellar models that are used throughout the \plato\ stellar analysis system (SAS). The grid is computed in two parts, an FGK-star grid and an M-dwarf grid. The FGK-star grid covers the mass range [0.6, 2.2]~\Msun\ and the metallicity range $[\mathrm{Fe/H}] = [-2.5, 0.5]$~dex. A star with a mass $\approx 2$~\Msun\ will be an A-type star while it is on the main sequence, but may evolve into an F-type subgiant star. The M-dwarf stellar model grid covers the mass range [0.1, 0.75]~\Msun\ and the metallicity range $[\mathrm{Fe/H}] = [-2.5, 0.3]$~dex. However, \plato\ will observe very few M-dwarfs with spectral types later than M4V because they are too faint. We therefore set the following limits for the selection of stars from DEBCat:
\begin{itemize}
 \item mass in the range [0.25, 2.2]~\Msun;
 \item \logg $> 3.5$;
 \item metallicity in the range [$-2.5$, +0.5]~dex;
 \item oblateness $\epsilon < 0.0025$.
\end{itemize}
Stars lacking metallicity or \teff\ estimates were not included. The equation we used to calculate the oblateness and the limit $\epsilon < 0.0025$ are the same as used by \citet{2025MNRAS.539.3741M} to define a sample of about 100 stars that were used to calibrate empirical relations to estimate the masses and radii of FGK-type stars. The limit on the star's oblateness is used because we want to exclude stars that are strongly affected by tidal forces in close, short-period binaries. We have included the estimates for \teff\ and $\log(L/L_{\odot})$ provided in DEBCat for stars selected from this catalogue. The DEBCat web page notes that ``{\it several slightly different effective temperature scales exist, and there are multiple alternatives for the luminosity of the Sun.}'' For this reason, we have not attempted to assign a benchmark level to these \teff\ and $\log(L/L_{\odot})$ estimates except in a few cases where we have been able to assign a benchmark level B1. 

In addition to stars selected from DEBCat \citep{southworth2015} we also collected individual targets with very accurate mass and radius, e.g.  $\alpha$~Cen~A and B \citep{kervella2017}. From the Araucaria project \citep{2023AA...672A.119G}, we selected 10 double-lined spectroscopic binaries for which high precision orbital parallaxes and accurate masses at an average level of 0.2\% were derived, including three that are also detached eclipsing binaries (AK~For, LL~Aqr, and V963~Cen).

We also included components of binary stars with accurate masses from the review on stellar masses by \citet{serenelli2021}. Components of bright visual binaries from their Table~3 are selected including the orbital parallaxes, and components from detached eclipsing binaries from their Table~2 are also included. Where possible we computed the rotational period from the measured projected rotational velocity $v\sin{i}$, as they are very often close to synchronization with the orbital period.

Finally, we specifically included M dwarfs in wide binaries, which have been extensively used in the literature to test analysis methods. The stars and most of their parameters were selected from \citet{Mann2015, Rains2021} and \citet{Olander2025a}. The magnitude range of this sample is $8 \lesssim G \lesssim 14$.

Masses are provided at benchmark level B3 and radii at level B2 for 14 of these stars, from empirical mass--luminosity relations in the K~band, and from SED-fitting methods, respectively \citep{Mann2015, Rains2021}.
Luminosities were calculated at benchmark level B1 for 16 stars from bolometric fluxes \citep{Mann2015, Rains2021} and \gaia\ parallaxes (Eq.~\ref{eqn:lumfromfbol}). Effective temperatures are provided at benchmark level B2 for all of these stars from spectroscopic analyses \citep{Mann2015, Rains2021, Olander2025a}. Surface gravities are provided for all stars, either from publications (\citealt{Rains2021}, from empirical mass-- and radius--luminosity relations in the K~band, 14 stars at level B3, and \citealt{Olander2025a}, from a fit of observed magnitudes to synthetic ones from evolution models, 9 stars at level B2), or calculated from their database masses and radii (Eq.~\ref{eqn:loggfrommasandrad}, 3 stars at level B3). For four of these stars we included ages determined from stellar evolution models by \citet{Mann2015}\footnote{We note that logarithmic values and uncertainties are given in the original publication, resulting in highly asymmetric linear uncertainties, with lower 1-sigma limits of 0.6 to 0.9 Gyr for the ages of these four stars, while the PBS database only quotes the larger uncertainties corresponding to the upper limits (see Sect.~\ref{sec:multiple}).}.

We note that the angular diameter (an ancillary parameter) provided in the database was calculated for the 14 stars which have radii from Eq.~(\ref{eqn:angfromrad}). See Sect.~\ref{sec:methods-Md} for further ancillary properties of this M-dwarf sample, such as metallicity, rotation, and activity. 

\subsection{Interferometry and astrometry}
\label{sec:interferometry}
For the definition of benchmark parameters dependent on angular diameters, we relied on interferometric and astrometric measurements. From the \gaia\ FGK benchmark stars (GBS, version 3, \citealt{soubiran2024}), we selected the stars spanning the main sequence and the subgiant branch. In that paper, the \teff, $L$, and $R$ were derived by compiling interferometric diameters from the JMDC catalogue \citep{chelli2016}, adhering to constraints on the quality and ancillary parameters as defined by \citet{salsi2020}. The angular diameters were then combined with \gaia\ DR3 parallax measurements \citep{gaiadr3}, extinction measurements from \cite{vergely2022} and bolometric flux measurements which were derived in the same work, in order to calculate \teff, $L$, and $R$. In addition, the mass and age were estimated by comparing these observed properties to several stellar models. The result is the addition of 105 stars with a median uncertainty $<1.5\%$ on the radius and $<2\%$ on the mass, where in particular for the mass we refer uniquely to its precision. 

A dedicated effort was made to obtain reference properties and parameters corresponding to the coolest \plato\ targets, the M~dwarfs encompassed by the \plato\ sample known as P4, in a similar way as for the GBS. The P4 sample is dominated by early- to mid-M dwarfs up to M4, and we selected benchmark M~dwarfs consistent with this spectral type range. We focused on stars for which interferometric measurements of the angular diameter have been published with an uncertainty of less than 5\%. Most of the angular diameter data were taken from \citet{boyajian2012}, \citet{Ellis2022}, and \citet{Rabus2019}, in order of decreasing number of stars, complemented by \citet{2014MNRAS.438.2413V,2011ApJ...729L..26V,2012ApJ...753..171V,2018ApJ...858...71S,2022AA...665A.120C}. Most of these stars have been used in other works for reference and validation \citep[e.g. ][]{2022ApJ...927..123S,2023A&A...673A.105B,2024MNRAS.529.3171R,Olander2025a,Olander2025b}. The magnitude range of this M-dwarf sample is $6 \lesssim G \lesssim 10$. We note that none of these stars are located in the (southern) \plato\ FOV.

Masses are provided at benchmark level B3 for 31 of these stars. For most of them the masses are based on empirical mass--luminosity relations in the K~band, and are taken from \citet{boyajian2012} or \citet{Rabus2019}, complemented by \citet{Mann2015}. In three cases, an empirical mass--radius relation was used \citep{2020AA...642A.115C,2022AA...665A.120C}. For one additional star, stellar evolutionary models were used \citep[GJ~393, level B2]{2018ApJ...858...71S}. Radii were calculated for all stars from their measured angular diameters and \gaia\ DR3 parallaxes (Eq.~\ref{eqn:radfromang}). Measurements of bolometric flux are provided for all stars, largely from the same sources as the angular diameter. However, for seven stars the bolometric flux was instead calculated from the luminosity values published in \citet{2020AA...642A.115C} and \citet{2022AA...665A.120C}, using \gaia\ DR3 parallaxes (Eq.~\ref{eqn:fbolfromlum}). For all stars with measured bolometric flux, the luminosity was in turn calculated (with \gaia\ DR3 parallaxes) from Eq.~(\ref{eqn:lumfromfbol}).

Ages are particularly difficult to determine for M~dwarfs from stellar evolution models, due to their long main-sequence life times. We include an isochrone age of 80(40)~Myr for the star GJ~393 from \citet{2018ApJ...858...71S} in the database, but we note that the authors quote another possible age range from 0.06 to more than 10~Gyr and argue that the star is likely an older metal-poor star. An alternative method is to assign an age to M~dwarfs from a stellar companion or a larger group of stars. Results from this age-by-association method were included for four of the interferometric M~dwarfs, from \citet{2022AA...665A.120C} for GJ~486 (based on a kinematics analysis by \citealt{2019A&A...627A.161P}), and from \citet{2023ApJ...954L..50E} for GJ~273, GJ~581, and GJ~699.

Effective temperatures are provided at benchmark level B1 for all of these stars, calculated from their measured angular diameters and bolometric fluxes (Eq.~\ref{eqn:tefffromangandfbol}). Surface gravities are provided at benchmark level B3 for 30 of these stars, calculated from a B1 radius and a B3 mass (Eq.~\ref{eqn:loggfrommasandrad}). One additional star has a B2 mass, and hence a calculated B2 \logg. For one more star a B2 \logg\ value was adopted from \citet{soubiran2024}. For ancillary properties of this M-dwarf sample in addition to angular diameter and bolometric flux, such as metallicity, rotation, and activity, see Sect.~\ref{sec:methods-Md}. 

\subsection{Asteroseismic samples}
Stellar properties can be highly constrained using global seismic parameters. We compiled \kepler\ FGK targets \citep{mathur2012,lund2017,creevey2017,silva_aguirre2017}. In the \gaia-\kepler\ Stellar Properties Catalogue \citep{berger2020} of almost 200\,000 stars, we selected 29 targets that fulfilled all the \plato\ requirements, except for the precision on metallicity that was slightly increased to $\Delta$[Fe/H]$\le0.12$ to have a sizeable sample. The stellar properties were derived from MIST isochrones \citep{choi2016}, broadband photometry, \gaia\ DR2 parallaxes and spectroscopic metallicities. The 29 selected stars all have a goodness-of-fit parameter larger than 0.99 indicating reliable ages. 

In the \apogee-\kepler\ catalogue \citep{claytor2020} of about 500 cool dwarf stars (\teff$<5000$~K, \logg$>4.0$), we selected 19 targets satisfying $\Delta M/M \le 5\%, \Delta R/R\le5\%$ and $\Delta \tau/\tau \le 10\%$. This cross-survey is one of the few providing a determination of the rotational period based on the analysis of \kepler\ light curves with wavelet decomposition. Also $\alpha$-element abundances are included. Uncertainties were not provided for \logg, so we computed them using the relations given in Appendix~\ref{app:params}. 

From the K2 asteroseismic KEYSTONE sample of dwarf and subgiant solar-like oscillators \citep[about 170 stars,]{2024AA...688A..13L}, we adopted the 33 stars from \citet{Lund2016} \citep[see also][]{Chaplin2015} that were observed during K2 Campaigns 1-3 and have stellar parameters from modelling of the global asteroseismic parameters \dnu and \numax.
We additionally included 31 stars from \citet{2024ApJS..271...17Z} who analysed \tess\ data.

\subsection{Large surveys}
\label{sec:large_surveys}
We also select stars from several spectroscopic samples, including large surveys. 
We selected the metal-poor dwarf stars Titans~I \citep{giribaldi2021}, and red giants Titans~II \citep{giribaldi2023} because of their precise and accurate \teff. This sample has \teff, \logg, and [Fe/H] derived by one different method each, therefore biases related to parameter intercorrelation are minimized. Effective temperature was derived by fitting the wings of the Balmer H$\alpha$ lines with 3D NLTE synthetic grids \citep{2018A&A...615A.139A} using the normalization method in \cite{giribaldi2019A&A...624A..10G}. The \teff\ scale of this sample was proven to be fully compatible with \teff\ inferred by interferometry of the \gaia\ FGK benchmark stars \citep{heiter2015AA...582A..49H} and with the \teff\ scale of the InfraRed Flux Method (IRFM) implementations of \cite{casagrande2010A&A...512A..54C,casagrande2021MNRAS.507.2684C}. The surface gravity of dwarfs was derived by isochrone fitting using Yonsei-Yale models \citep{kim2002, yi2003} and \gaia\ parallaxes. 

In the SPOCS catalogue \citep{valenti2005} of about 1000 FGK dwarf stars, we selected nine bright targets. Luminosities are computed from apparent magnitudes, parallaxes and bolometric corrections, radii from luminosities and spectroscopic effective temperatures, and masses from spectroscopic \logg\ and radii. Ages are derived by isochrone fitting, with a conservative 10\% uncertainty chosen when reported as zero. 

Finally, the golden sample of \gaia\ astrophysical parameters \citep{2023AA...674A..39G} provides very high precision parameters. We selected among this catalogue those FGK dwarfs and subgiants in the \plato\ fields of view with $G < 9.1$, and separated the stars in bins of \teff\ and \logg\ to select the 1--3 brightest targets in each bin. This resulted in about 60 FGK stars. 

Furthermore, we included 40 M~dwarfs with $13<G<16$. The parameters provided for this sample in the present database are radius, luminosity, \teff, and \logg\ for all stars, masses for a subset, and ages for the FGK part only.

We initially considered including targets from the \galah\ DR3 survey \citep{2021yCat..75060150B} and the extended \hipparcos\ compilation \citep[XHIP,][]{anderson2012}. However, we ultimately decided to exclude both catalogues from the final database. While large spectroscopic surveys like \galah,  \apogee, or \gaia-ESO provide valuable atmospheric parameters (\teff, \logg, and [Fe/H]), they do not directly yield the model-independent fundamental masses and radii required for our benchmark scale. Potential systematic trends in \teff\ and \logg\ currently limit their immediate applicability as high-fidelity reference standards \citep[e.g.][]{2021yCat..75060150B}. In the case of XHIP, the available age estimates are based on old data and original \hipparcos\ parallaxes. Consequently, they were deemed not useful as modern benchmark values to test age estimates that are based on highly precise \gaia\ parallaxes and photometry.

\subsection{Solar analogues}
We explicitly compile individual, well-studied solar analogues in the present catalogue. These include prime standards (i.e. solar twins) such as 18~Sco \citep{bazot2011} and HIP~56948 \citep{melendez2012}. In addition to its precisely determined fundamental parameters, 18 Sco has a well-characterised rotation period and magnetic activity cycles. Long-term observations reveal a shorter chromospheric activity cycle of approximately 7~yr and a longer periodicity of approximately 15~yr, interpreted as a Hale-like magnetic cycle associated with changes in the star’s global magnetic geometry \citep{2023ApJ...958...57D}. This makes 18~Sco a particularly valuable benchmark not only for fundamental stellar parameters, but also for the rotational and magnetic evolution of solar-type stars. We also include the solar analogue components from binary systems where asteroseismic and transit constraints are available (such as LL~Aqr~A \citep{graczyk2016} Rigil Kentaurus \citep[$\alpha$~Cen~A,][]{morel2018} and 16~Cyg~B \citep{morel2021}.

In addition, 28 solar-analogue stars were selected from a parent table of $\sim$5\,000 solar analogues compiled by \citet{2023AA...674A..39G}. These targets were chosen to have parameters closest to solar values, satisfying similarity thresholds of $\pm$100~K in \teff, $\pm$0.25~dex in \logg, $\pm$0.22~dex in \feh, $\pm$0.05~\Msun\ in mass, and $\pm$0.2~\Rsun\ in radius, while also passing a rigorous filter on \gaia\ astrometric, photometric, and processing quality flags. The selected stars represent the brightest solar analogues in the parent sample ($7.2 < G < 10.0$) and all have published RVS spectra.

\section{Methods}
\label{sec:methods}

\subsection{Nomenclature}
The nomenclature used in the PBS database is designed to systematically identify each star with a \emph{main} name following a prioritized scheme plus an \emph{identifier}. This will ensure consistency and ease of cross-identification. 

The \emph{main} name of a star is chosen to comply with the \simbad\ database conventions. It is selected among commonly used designations that allow a reliable match with major astronomical catalogues. 
The priority for selecting the \emph{main} name follows a hierarchical order based on the star's common catalogue identifiers, beginning with Bayer or Flamsteed designations (\emph{e.g.}, \texttt{alf~Cen}\footnote{We use fixed-width font to emphasize where identifiers correspond to an exact text string in our database or in the \simbad\ database} or \texttt{18~Sco}), followed by cluster designations (\emph{e.g.}, \texttt{Cl* NGC 6809 SAW V54}), variable star names (from GCVS, \emph{e.g.}, \texttt{AI~Phe}), and then other catalogues such as HD designation (Henry Draper, \emph{e.g.} \texttt{HD 179260}), EBLM designation (Eclipsing Binary with a Low Mass star, \emph{e.g.} \texttt{EBLM J0608-59}), TYC designation (Tycho, \emph{e.g.} \texttt{TYC 8533-950-1}), and 2MASS (Two Micron All Sky Survey, \emph{e.g.} \texttt{2MASS J03553688+5214291}). No use of TIC, KIC or BD appear in the \textit{main} name. In addition, 13 benchmark stars have only a \gaia~DR3 \emph{main} name. Using this \emph{main} name, all of these stars are retrievable in \simbad.

Each star entry is associated with a unique \emph{identifier} from either the \gaia\ DR3 or \hipparcos\ catalogue. We prioritize \gaia\ DR3 identifiers due to their recent and comprehensive astrometric coverage, falling back to \hipparcos\ identifiers for bright stars ($G \lesssim 3$) where \gaia\ sources are unavailable. In total, the sample includes 972 unique \gaia\ DR3 identifiers and 7 unique \hipparcos\ identifiers. Within the \gaia\ DR3 subset, 123 sources are unresolved binaries and two are unresolved triples: \texttt{Gaia DR3 3424609014036545920} (\texttt{* 1 Gem} A, Ba, Bb) and \texttt{Gaia DR3 3312892757136810880} (\texttt{HD 28363A}, Ba, Bb in the Hyades). Among the 7 \hipparcos\ identifiers, one is an unresolved system: \texttt{HIP 24608} ($\alpha$~Aur A and B).

For some bright and nearby binary systems the two components of a binary star system have different \gaia\ DR3  identifiers and separate entries in the \simbad\ database. In these cases we include the component identifier as part of the \emph{main} name, e.g. \texttt{16~Cyg~A} and \texttt{16~Cyg~B}. When the two components of a binary are not resolved in \simbad, then the \texttt{component} field in the PBS database is used to distinguish them. Magnitudes refer to the magnitude of the source, not the magnitude of the individual components for unresolved sources, e.g. eclipsing binary stars. Apparent magnitudes for all sources are taken from the \simbad\ database. We use the \gaia\ G magnitude when available from \simbad, otherwise the V magnitude is used.

\subsection{Homogenisation of parallaxes} 
\label{sec:parallaxes}
As we aim to provide a consolidated catalogue of stellar parameters, we re-calculate the luminosity and radius using a consistent approach for the distance determination.
Unless indicated in the database, we exploited the parallaxes from \gaia\ DR3 and applied the zero-point correction of \citet{lindegren2021}, including for sources that have astrometric solutions in the non-single star tables, as suggested by \citet{2024PASP..136i4203N}. The cases where we did not adopt this approach was when the \hipparcos\ parallax was more precise, or for some specific binary systems. Of the 1094 stars in the catalogue with a parallax, the \gaia\ DR3 parallaxes are used for 1003.

\subsection{Handling multiple values from various sources and uncertainties}
\label{sec:multiple}
As a catalogue of benchmark values, we decided to provide only one value per stellar property per star.
Between 100-150 benchmark stars appeared with multiple values for a given stellar property from different sources in the original compilation of the catalogue. To disambiguate this, we proceeded as follows: in the cases where the various values were in agreement with each other, we either selected the most recent reference, or the reference with the largest provided number of properties, or the reference with the best precision. The reference leading to a stellar property with a B1 value was always the preferred one. 

For compiled literature parameters, we adopt published symmetric uncertainties (taking the larger absolute value if asymmetric). For properties derived within our pipeline, uncertainties are propagated in quadrature using the even-numbered equations in Appendix~\ref{app:params}, assuming uncorrelated input observables. This is well-justified as our primary observables (\gaia\ DR3 parallaxes, angular diameters, and bolometric fluxes) derive from independent techniques and instruments. While Monte Carlo sampling \citep[e.g.,][]{stassun2019} is optimal for jointly derived parameters, analytical quadrature provides a conservative, computationally efficient propagation for independent observables.

In the cases where the various values of a stellar property were not in agreement, we tried to understand why and selected the value that was considered to be more consistent with other stellar properties. 
The most striking example concerns $\alpha$~Cen~B (main name \texttt{*~alf02~Cen}, identifier HIP~71681), where a discrepancy exceeding $11\sigma$ (an absolute difference of $0.0532$~\Msun) exists between the selected B1 mass of $0.9092\pm0.0025$~\Msun\ \citep{2021AJ....162...14A}, obtained from a joint astrometric and SB2 solution, and the B2 mass of $0.856\pm0.004$~\Msun\ \citep{2024AA...684A...9S}, estimated from stellar evolution models. 
Another notable case is the radius of Procyon~A (main name \texttt{*~alf~CMi~A}, identifier \texttt{HIP~37279}). Its B2 radius of $1.919\pm0.015$~\Rsun\ \citep[from the SPOCS sample;][]{valenti2005} differs by $7\sigma$ ($0.1236$~\Rsun) from the calculated B1 radius of $2.0426\pm0.0093$~\Rsun, derived from its angular diameter \citep{2021AJ....162..198B} and \hipparcos\ parallax \citep{van_leeuwen2007} using Eq.~(\ref{eqn:radfromang}). 
Finally, for the luminosity of Procyon~A, the value of $6.08\pm0.19$~\Lsun\ derived from \teff\ and $R$ via the Stefan-Boltzmann law (Eq.~\ref{eqn:lumfromteff}) deviates by nearly $5\sigma$ (an absolute difference of $0.969$~\Lsun) from the selected B1 value of $7.049\pm0.064$~\Lsun\ computed directly from the bolometric flux \citep{soubiran2024} and parallax (Eq.~\ref{eqn:lumfromfbol}).
In all cases we aimed to trace the source of a property to the original work in which that property was measured or estimated and include the corresponding publication as a reference.

\subsection{Mean stellar density}
The value of $\log\rhostar$ and its uncertainty were computed directly from the mass and radius for benchmark stars where both these quantities are present in the database. In this case the lowest benchmark level from these input quantities is inherited.
We also include a sample of benchmark stars in transiting exoplanet systems and eclipsing binaries with low mass companions (EBLM systems) where \rhostar\ has been computed from the transit light curve, as described in Sect.~\ref{sec:rhostar}. 
The data for these stars are all taken from \citet{2025MNRAS.539.3741M}. 
These stars can also be used as benchmark stars for \logg\ estimates. 
Of the 36 stars from \citet{2025MNRAS.539.3741M}, 28 are the primary stars in EBLM systems with a companion that is at least 8 magnitudes fainter than the primary star in the V band. 
EBLM systems can be identified by the presence of an entry for the quantity $\frac{d\log(\rho/\rho_{\odot})}{dM/M_{\odot}}$ in the database. 
This quantity can be used to apply a small correction to these values of \rhostar\  or \logg\ in cases where an independent estimate of the stellar mass is preferred to the value given in the database. 
For the binary systems EBLM~J0113+31 and EBLM~J0608$-$59, the value of $\frac{d\log(\rho/\rho_{\odot})}{dM/M_{\odot}}$ is given as 0.000 because the mass ratio has been measured directly for these binary systems \citep{2022MNRAS.513.6042M, 2024MNRAS.531.4577M}. 
The values of mass, radius, \rhostar\  and \logg\ are therefore benchmark level B1, as is the effective temperature of these stars measured from their radius, parallax and bolometric flux. 
The apparent magnitudes and flux ratios at several optical and near-infrared wavelengths for these EBs are provided in \citet{2022MNRAS.513.6042M,2024MNRAS.531.4577M} and \citet{2023MNRAS.522.2683M}.

\subsection{Limb darkening}
Values of the limb-darkening parameters ($h^{\prime}_1,h^{\prime}_2$) for FGK stars with transiting exoplanets in the \kepler\ and \tess\ bands are taken from \citet{2023MNRAS.519.3723M}.
These and the values of ($h^{\prime}_1,h^{\prime}_2$) measured using very similar methodology based on \tess\ light curves of solar-type stars transited by very low mass stars by \cite{2025MNRAS.537.3943F} are considered to be benchmark level B1. Values of ($h^{\prime}_1,h^{\prime}_2$) for the sample of stars studied by \citet{2023MNRAS.519.3723M} have also been measured by \citet{2024MNRAS.534.3893V} using a regularisation technique to improve the precision of these measurements. We decided to use the measurements from \citet{2023MNRAS.519.3723M} in our table of benchmark values so that all the limb darkening values are measured in a consistent way. 

\subsection{Calculations/update of stellar properties}
The derivation and homogenisation of fundamental stellar properties were performed using \texttt{Python} processing pipeline developed specifically for this work package (\texttt{benchpy}).
This automated pipeline systematically evaluates the available input parameters (such as parallax, angular diameter, bolometric flux, and mass) to determine the optimal mathematical relations for computing missing properties or updating existing ones. Specifically, \texttt{benchpy} computes the stellar radius, surface gravity, effective temperature, luminosity, and empirical mass by applying the physical relations detailed in Appendix~\ref{app:params}. 

To ensure physical consistency, the software robustly propagates uncertainties from the input observables to the derived quantities assuming uncorrelated errors among independent observables, while enforcing rigorous validity checks against predefined physical limits. In instances where multiple derivation paths are available, the pipeline resolves degeneracies by preferentially selecting the method that relies on the most direct observables, thereby yielding the highest possible benchmark level for the final catalogue of stellar parameters.
  
We performed internal consistency checks to verify the mathematical self-consistency of the catalogue across defining equations sharing common observables.
Specifically, we verified the consistency between the reference $\log g$ values compiled in the catalogue (derived predominantly from independent spectroscopic or asteroseismic analyses) and the surface gravities calculated directly from the assessed stellar masses and radii via Eq.~\ref{eqn:loggfrommasandrad}. For the $\sim800$ stars with available masses, radii, and reference $\log g$, the mean difference is under $0.001$\,dex (standard deviation $0.030$\,dex). 
Over 90\% of stars agree to within $0.05\,\text{dex}$, and 97\% agree to within $0.10\,\text{dex}$.
We also verified the mutual consistency between stellar radii, luminosities, bolometric fluxes, and parallaxes. For the 850 stars with all required quantities available, the calculated luminosities, Eq.~(\ref{eqn:lumfromfbol}), and bolometric fluxes, Eq.~(\ref{eqn:fbolfromlum}), show median relative differences under $0.01\%$ (standard deviations of $2.1\%$ and $1.7\%$, respectively), with over $97\%$ of stars agreeing to within $1\%$.

\subsection{Rotation periods}
For stars in detached eclipsing binaries and bright visual binaries, estimates of the rotation period, $P_\mathrm{rot}$, have been obtained through various methods. In a number of cases, $P_\mathrm{rot}$ could be adopted directly from the source publications. This includes measurements from the out-of-eclipse portion of the light curve, analysed via autocorrelation or periodogram approaches \citep[e.g.,][]{2017ApJ...845...72K}, as well as individual determinations using the Rossiter-McLaughlin effect \citep{1924ApJ....60...15R, 2018MNRAS.478.1942S}. Furthermore, when authors explicitly established tidal synchronisation, we adopted $P_\mathrm{rot} = P_\mathrm{orb}$ along with its corresponding uncertainties. 

However, in most cases, the rotation period estimates were computed from the projected equatorial rotation velocity, $v\sin i_\text{rot}$, measured from spectral line broadening and converted to $P_\mathrm{rot}$ using the direct radius measurement. 
These estimates rely on the assumption of spin-orbit alignment, meaning the rotation axis of the star, $i_\mathrm{rot}$, is assumed to be aligned with the binary star orbit ($i_\mathrm{rot} = i_\mathrm{orb}$). For detached eclipsing binary stars, this orbital inclination is typically $i_\text{orb} \approx 90^\circ$. This assumption has been tested for a small sample of solar-type stars with M-dwarf companions and found to be approximately correct for orbital periods up to about 15 days \citep{2026MNRAS.550g1334S}. It is not yet fully known under what circumstances this approximation breaks down for late-type stars, though it is known to be a poor approximation for some early-type stars in short-period binary systems \citep{2022ApJ...933..227M} and F-type stars in very eccentric orbits with $P > 50$~d \citep{2024ApJ...975..149M}. Assuming $i_{\rm rot} = 90^\circ$ yields an estimated period $P_{\rm est} = 2\pi R / (v \sin i_{\rm rot})$, which represents an upper limit to the true period ($P_{\rm rot} = P_{\rm est} \sin i_{\rm rot} \le P_{\rm est}$). In the database, these derived values are explicitly flagged as upper limits via \texttt{prot\_flag}, while \texttt{prot\_type} distinguishes direct photometric modulation, $v \sin i$ estimates, and tidally synchronised orbital periods.

For single stars and wide binary components lacking photometric periods, $v \sin i$ values are compiled from high-resolution spectroscopic literature. However, for slow rotators, measurements approach instrumental resolution limits and are strongly degenerate with macroturbulence broadening and the unknown inclination angle $\sin i$. In this regime, derived $P_{\rm rot}$ estimates carry large uncertainties and generally represent upper limits. For information on rotation periods of M~dwarfs see Sect.~\ref{sec:methods-Md}. Additionally, photometrically derived modulation periods trace active region latitudes; surface differential rotation and magnetic cycle migration can thus introduce small temporal variations into the measured $P_{\rm rot}$.

\begin{figure}
    \centering
    \includegraphics[width=\linewidth]{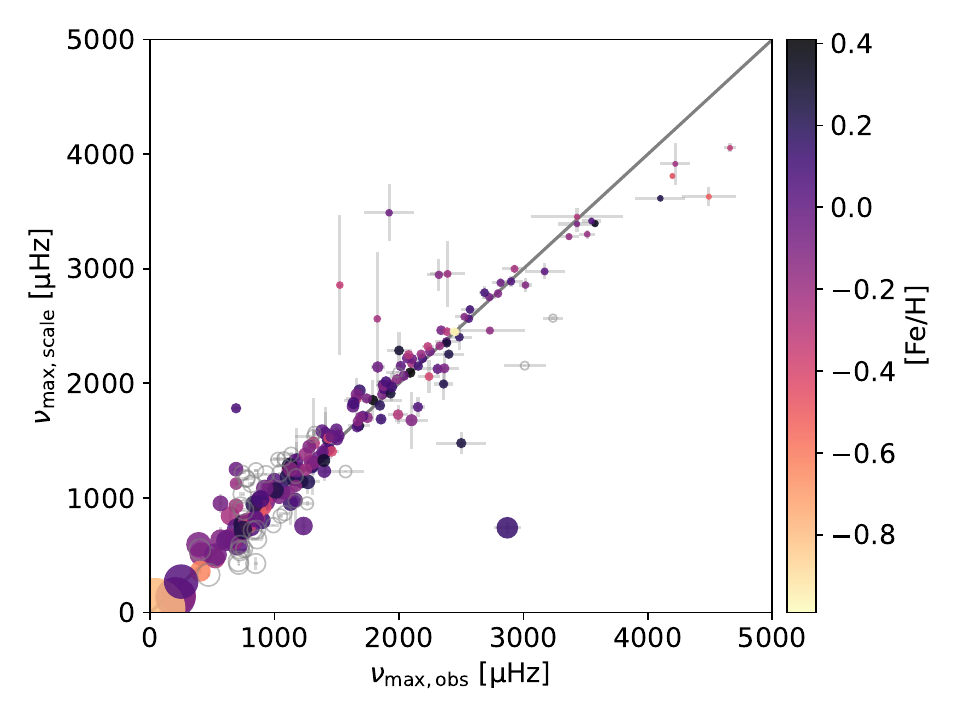}
    \caption{Comparison of the global asteroseismic frequencies at maximum power from scaling relations for stars in the \plato\ benchmark star database to published estimates from various catalogues, colour-coded by metallicity. Open symbols have no metallicity. Symbol size is proportional to radius. Stars with a standard error larger than 0.1 for \logg\ were excluded from the figure.}
    \label{fig:astero}
\end{figure}

\subsection{Global asteroseismic parameters}
The frequency at maximum power, $\nu_{\text{max}}$, and the large frequency separation, $\Delta\nu$, have been compiled from several catalogues. In order of preference, we adopted the following values: 26 values for various targets from \citet{soubiran2024}; for \kepler\ targets, one value from \citet{2022AA...657A..31M}, 75 values from \citet{2017ApJS..233...23S}, 12 values from \citet{lund2017}, and 30 values from \citet{2016MNRAS.456.2183D}; for \tess\ targets, 45 values from \citet{2025AA...701A.285L}, 31 values from \citet{2024ApJS..271...17Z}, and four values from \citet{2023AA...669A..67H}; and finally, 165 values for K2 targets from \citet{2024AA...688A..13L}.

The values of $\nu_{\text{max}}$ and $\Delta\nu$ from these catalogues are compared to the predicted values in Fig.~\ref{fig:astero}, using the following scaling relations:
\begin{equation}
\label{eq:numax}
    \frac{\nu_{\text{max}}}{\nu_{\text{max},\odot}} = \left(\frac{g}{g_\odot}\right) \sqrt{\frac{T_{\text{eff}}}{T_{\text{eff},\odot}}}
\end{equation}
assuming solar values of $\nu_{\text{max},\odot}$ and $\Delta\nu_\odot$ given in Table~\ref{tab:ref} from \citet{2011ApJ...743..143H}. 
\citet{hekker2020} discusses scaling relations not only in the context of dwarfs but also of subdwarfs and giants.
We note a systematic deviation at high frequencies ($\nu_{\text{max}} \gtrsim 2500\ \mu$Hz), which corresponds to the unevolved main-sequence dwarf regime. This skewness might represent a known systematic limitation of the classical scaling relation, which assumes a simplified isothermal atmosphere and neglects variations in the mean molecular weight, the first adiabatic exponent, and the outer boundary structures of the atmospheres of dwarf stars \citep{2017ApJ...843...11V, 2024ApJ...962..118Z}.

\subsection{Ancillary properties for M dwarfs}
\label{sec:methods-Md}
The PBS database includes two samples of early- to mid-M~dwarfs, in wide binary systems and with precise angular diameters from interferometry (see Sects.~\ref{sec:binary} and ~\ref{sec:interferometry}, respectively, and Table~\ref{tab:data_compilation_summary}). In addition to the data described in those sections, we provide metallicities, rotation periods, and an indicator for chromospheric activity for these samples. The metallicities of about half of the M~dwarfs in the the wide-binary sample are those resulting from an analysis of optical high-resolution spectra of the respective FGK primaries by \citet{2008AA...487..373S,2016ApJS..225...32B,montes2018,2020ApJ...898..119R}.
For the other M~dwarfs in that sample, the metallicities are based on an analysis of high-resolution near-infrared spectra \citep[][9 stars]{Olander2025a}, or a fit of low-resolution flux-calibrated spectra \citep[][3 stars]{Mann2015}. Metallicity values are also provided for all M~dwarfs in the interferometry sample, and were taken from \citet{2021AA...656A.162M} 
for the majority, supplemented by \citet{Mann2015} and \citet{2020AA...644A..68M}. 

Rotation periods were included in the database for 27 stars (26 in the interferometry sample and the wide binary component HD~154363B), taken from \citet{2015MNRAS.452.2745S, 2016AA...595A..12S, 2018AA...612A..89S, 2019AA...621A.126D, 2019AA...625A..17D, 2022AA...665A.120C, 2023ApJ...954L..50E, 2024AA...684A...9S}. All of these were determined from an analysis of the modulation of a photometric or spectroscopic time series by magnetically active regions.

\begin{figure}
    \centering
    \includegraphics[width=\linewidth]{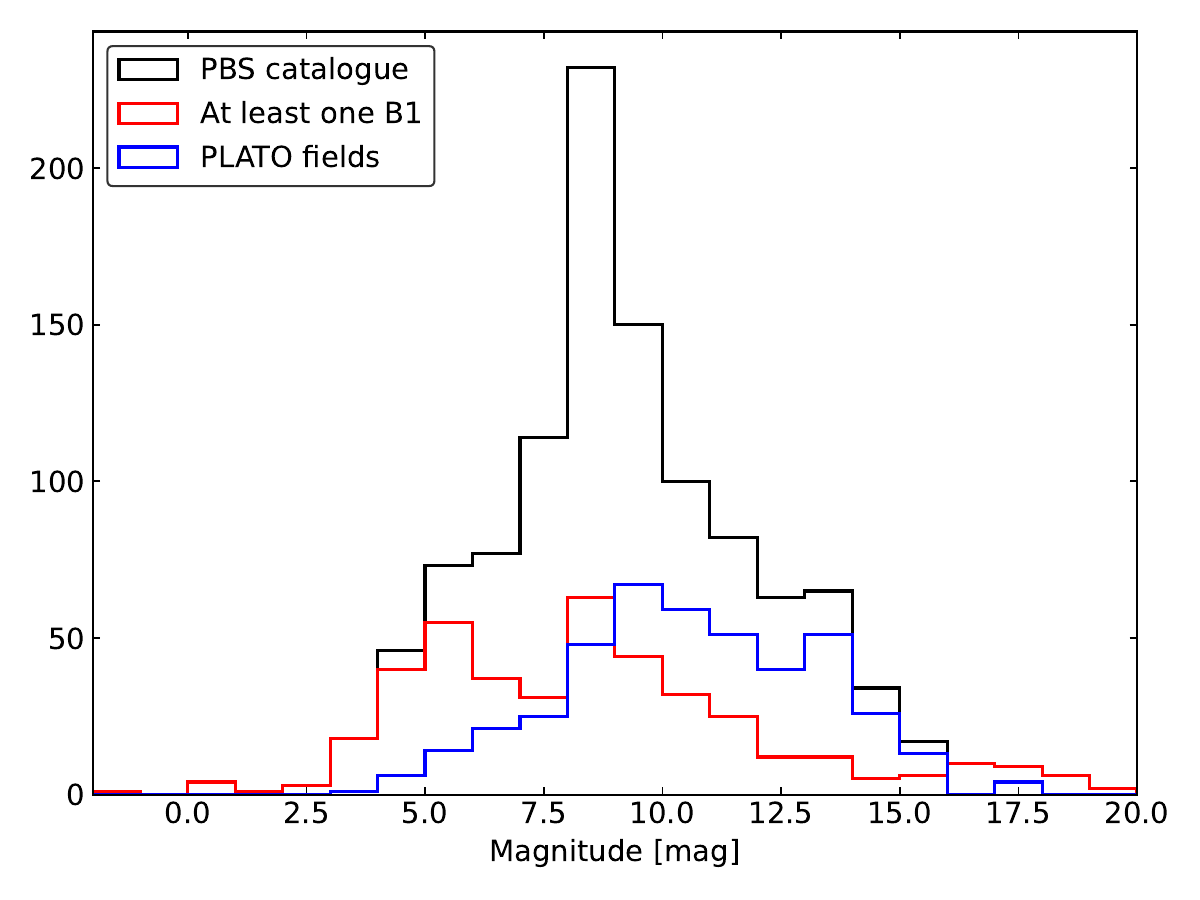}
    \caption{Distribution of \gaia\ $G$ magnitudes for different samples of the \plato\ benchmark stars catalogue. Magnitudes lower than 3 are in $V$ band.}
    \label{fig:histo_mag}
\end{figure}

M~dwarfs often show signs of chromospheric activity and significant magnetic fields, which can have an impact on the measurement or interpretation of stellar parameters. In order to assess the level of activity of the stars in the two M~dwarf samples, we compiled X-ray flux measurements from the literature, complemented by data obtained directly from the XMM-Newton \citep[4XMM-DR14,][]{2020A&A...641A.136W}, ROSAT \citep[2RXP or 2RXS,][]{2000yCat.9030....0R,2016A&A...588A.103B}, or eROSITA \citep[eRASS1,][]{2024A&A...682A..34M} catalogues. X-ray fluxes are available for 13 stars in the wide-binary sample and 38 stars in the interferometry sample. We stress that these are to be used as ancillary information for a rough assessment of stellar activity only. The X-ray fluxes are not on a homogeneous scale, their quality has not been investigated, and they are therefore not intended for benchmarking. Since the X-ray properties are only available for a small number of stars among the PBS they are listed separately in Table~\ref{tab:mdwarfs_xray}.

The quantity listed in the table is the logarithmic X-ray to bolometric flux ratio $\log(f_X/f_{\mathrm{bol}})$, which is equivalent to the X-ray to bolometric luminosity ratio $\log(L_X/L)$. Here, $f_X$ or $L_X$ are the X-ray surface flux or luminosity in the 0.1--2.4 keV band, and $f_{\mathrm{bol}}$ or $L$ are the bolometric flux or the (bolometric) luminosity. Data ($L_X/L$) were taken from \citet{2025A&A...694A..93Z} for 25 stars (who used X-ray data from ROSAT, eROSITA, XMM-Newton, or Chandra, and luminosities from \gaia\ DR3), from \citet{2022A&A...662A..41R} for 9 stars (who used X-ray data from ROSAT 1RXS and luminosities from \citealt{2019A&A...625A..68S}), and from \citet{2023A&A...676A..14C} for the binary components GJ~745 A and B (based on XMM-Newton data). For the remaining 15 stars we extracted X-ray count rates from XMM-Newton using the web interface provided by the XMM-Newton Survey Science Centre\footnote{\url{https://xmm-catalog.irap.omp.eu/}, accessed 2025-06-11}  \citep{2017ApJ...839..125Z}, and from ROSAT 2RXP, eRASS1, and ROSAT 2RXS via VizieR. The X-ray count rates were converted to X-ray fluxes using the instrument-specific conversion factors given in Table~1 of \citet{2023A&A...676A..14C}. The X-ray fluxes were then divided by the bolometric fluxes from the PBS database to obtain the X-ray to bolometric flux ratio.

One star in the interferometry sample is currently lacking an X-ray flux: HD 173739 (GJ 725A). The 13 stars in the wide-binary sample without X-ray fluxes are the following: 2MASS~J03553688+5214291, 2MASS~J03044335+6144097, 2MASS~J03150093+0103083, 2MASS~J14045583+0157230, HD~11964~B, TYC~5748-175-1, TYC~6373-228-1, HD~46375~B, HD~164595~B, 2MASS J01542093-1543479, HD~38529~B, HD~18143~C, HD~183870~B.

\subsection{Cluster membership}
Probable or possible member stars of open clusters were identified using the catalogue of \citet{2023AA...673A.114H}. 
Cross-matching with this catalogue was performed using the \gaia~DR3 identifier. 
We note that some of the stars in the PBS database have identifiers in the \simbad\ database starting with \texttt{Cl*}, indicating cluster membership, but they do not appear in the catalogue by \citeauthor{2023AA...673A.114H}.
This includes members of well-known open clusters such as the Pleiades (\texttt{Cl* Melotte 22 \dots}), and the Hyades (\texttt{Cl* Melotte 25 \dots}).
A few eclipsing binaries in the PBS database are members of globular clusters.
These eclipsing binaries in globular clusters can be identified from their names, e.g. \texttt{Cl* NGC 6121 \dots} (M4), NGC~104 (47~Tuc), etc. 

\begin{figure*}
\centering
\includegraphics[width=\linewidth]{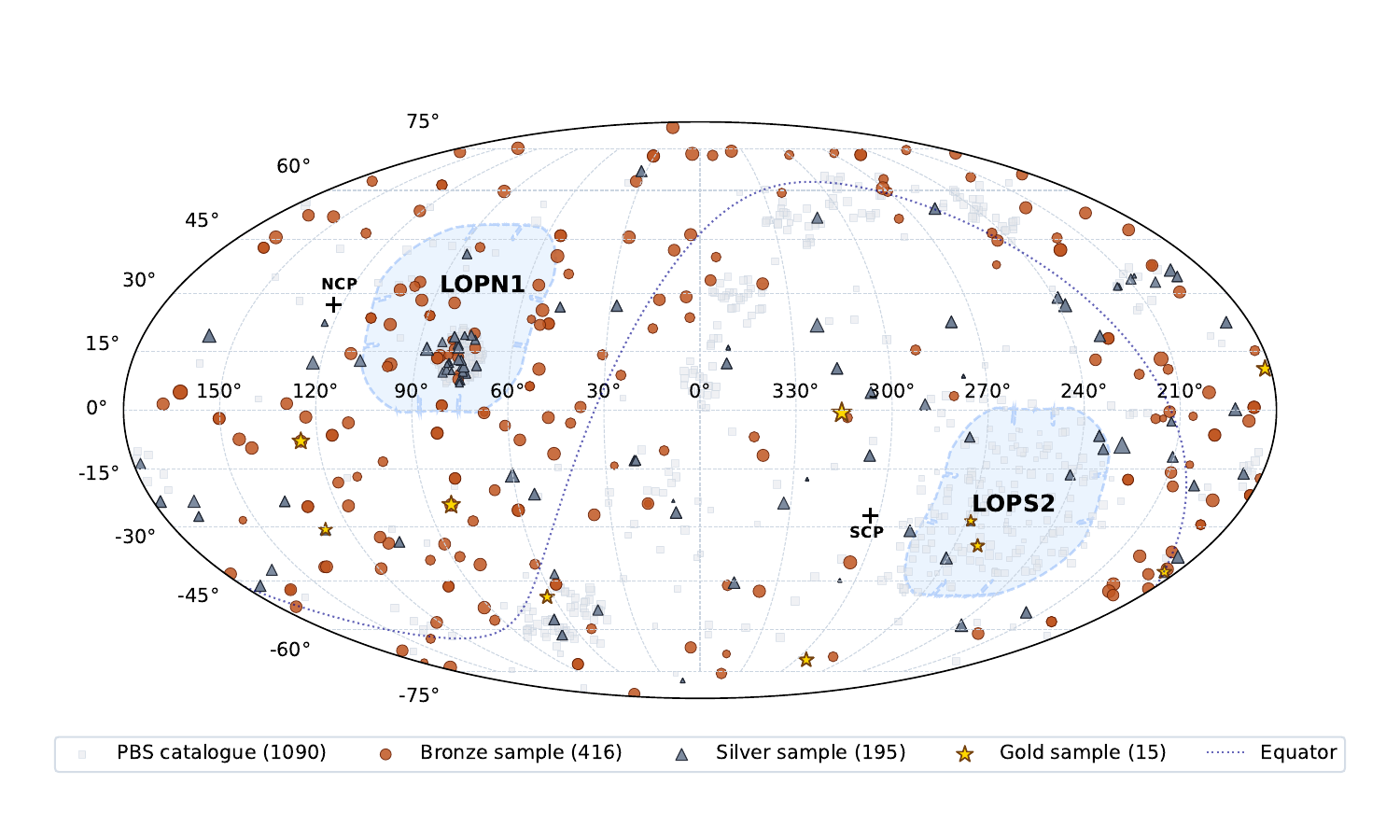}
\caption{Sky map of the \plato\ benchmark stars in Galactic coordinates. Different symbols indicate benchmark quality samples as defined in the text (Sect.~\ref{sec:discussion}). Symbol size increases with apparent brightness.}
\label{fig:skymap}
\end{figure*}

\begin{figure*}
\centering
    \includegraphics[width=\linewidth]{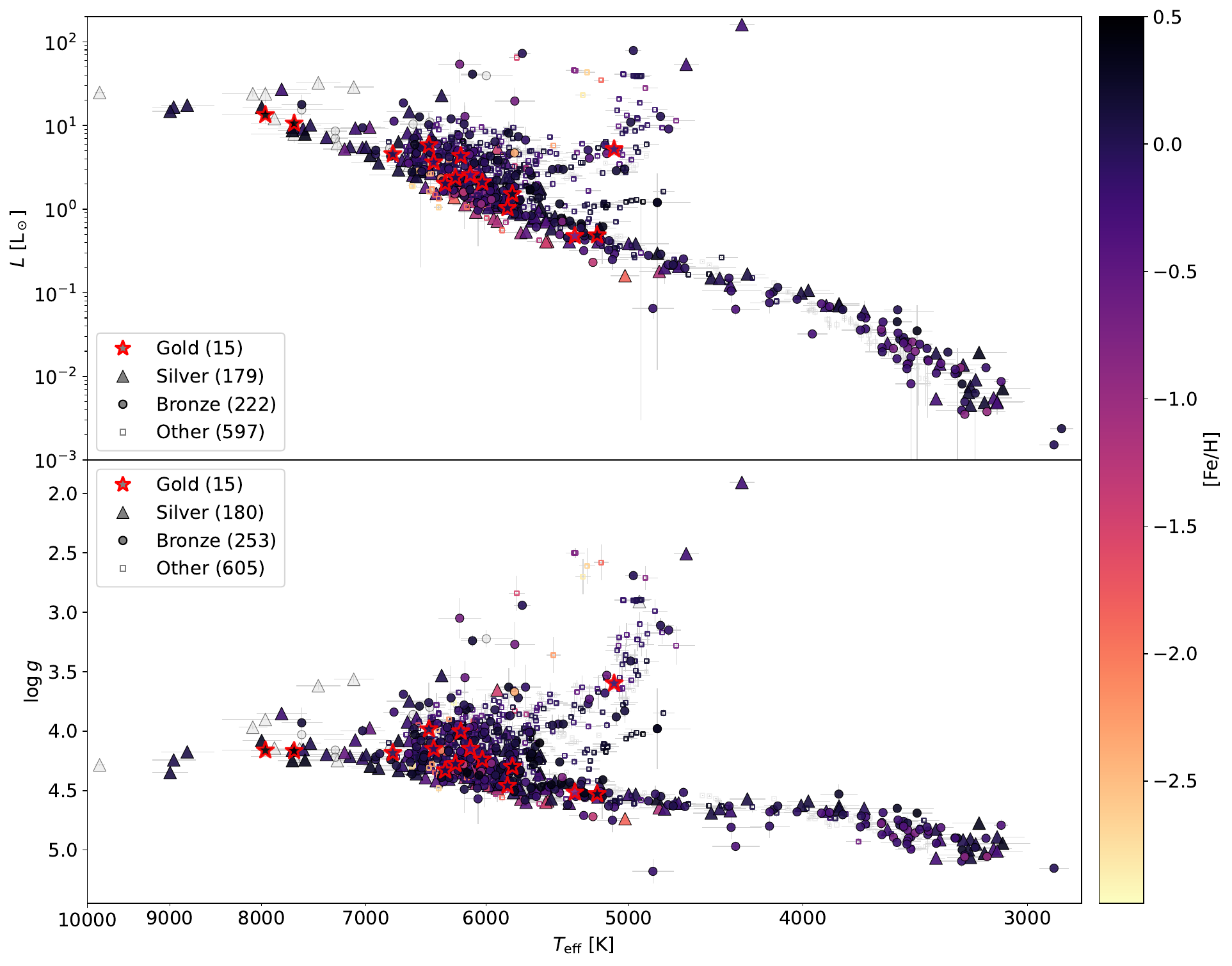}
    \caption{Hertzsprung-Russell (top) and Kiel (bottom) diagrams. Custom symbols distinguish the benchmark quality: large red-bordered stars (\emph{gold}), medium filled triangles (\emph{silver}), small filled circles (\emph{bronze}), and tiny open squares for the remaining stars. They are colour-coded by metallicity. Empty symbols means that no metallicity values are present in the database.}
    \label{fig:hrd_kiel}
\end{figure*}

\section{The \plato\ benchmark stars catalogue}
\label{sec:cat}

The \plato\ benchmark stars catalogue (PBS, v1.0) is accessible via CDS VizieR and features a dedicated web portal\footnote{\url{https://astro.ulb.ac.be/pbs}}. To optimise the discoverability of benchmark targets, the portal includes a unified search field with a fallback to \simbad. Users can browse the database and use column filtering to refine custom samples. Furthermore, a TAP service supporting ADQL queries is available for more complex and customisable requests. The website also allows users to visualise the entire benchmark sample, or specific subsamples, through interactive sky maps, Hertzsprung-Russell (HR) diagrams, and mass-luminosity as well as mass-radius relations.

\subsection{General properties}
\label{sec:cat-gen}
This first version of the PBS database gathers 1090 benchmark stars covering magnitudes from V $= -1.5$ (Sirius) to G $=$ 19.8 (an eclipsing binary in the globular cluster 47 Tuc, \texttt{\gaia~DR3~4689641564666423040}), see Fig.~\ref{fig:histo_mag}. The median $G$ magnitude of the whole PBS catalogue is 8.9 while slightly less bright in the \plato\ fields (10.6), and slightly brighter for stars with at least one B1 property (8.3). 
Eight stellar properties are assessed into three benchmark levels (B1, B2, and B3), reflecting their reliability as defined in Sect.~\ref{sec:properties}. Table~\ref{tab:parameter_summary} summarizes the distribution over benchmark levels of the eight stellar properties assessed. 
Out of the 1090 stars in the database, surface gravity (\logg) is the most widely assessed parameter with 982 stars, followed by mass ($M$, 840 stars) and effective temperature (\teff, 831 stars).  496 stars have B2 age in the catalogue  and limb darkening coefficients are reported for 62 stars.
Fig.~\ref{fig:skymap} displays a Mollweide projection of the Galactic celestial distribution of the PBS catalogue. About 285 benchmark stars are located in the \plato\ Northern Long-duration Observation Phase field (LOPN1; $\alpha \simeq 277.1^\circ$, $\delta \simeq 52.9^\circ$, \citealt{2022A&A...658A..31N} and priv. comm.), while 143 stars are in the Southern field (LOPS2; $\alpha \simeq 95.3^\circ$, $\delta \simeq -47.9^\circ$ \citealt{nascimbeni2025}). 
The high-density northern cluster in LOPN1 corresponds to the original \textit{Kepler} field, and the ecliptic bands reflect the K2 campaigns. In contrast, the concentration of 143 stars in LOPS2 is a deliberate selection rather than a natural overdensity. Because LOPS2 is \plato's first long-pointing field, we systematically cross-matched candidates within its footprint to support early in-flight pipeline validation \citep[scvPIC;][]{zwintz2026, nascimbeni2025}. The details of the cross matches are given in Table~\ref{tab:plato_fov_crossmatch}.
The distributions of assessed values per benchmark level (excluding mean density and limb-darkening coefficient) are displayed in Figure~\ref{app:histo}.

\subsection{Fulfilment of work package requirements}
\label{sec:fullfilment_requirements}
We compared the initial contents of the PBS catalogue with the pre-defined scientific requirements gathered from the various \plato\ stellar science work packages (see Sect.~\ref{sec:requirements}). Several key requirements are already successfully met; for example, the selected M dwarfs with precise \teff, \logg, and \feh\ are highly suitable for validating 3D model atmospheres, provided high-resolution, high-signal-to-noise near-infrared spectra are acquired to measure line asymmetries. In other cases, the PBS sample has bypassed original observational limitations: while early requirements anticipated that high-precision limb-darkening profiles could only be measured via long-baseline interferometry on very bright stars ($\theta > 0.8$\,mas), the database includes B1-level limb-darkening coefficients derived much more precisely from space-based transit light curves of hot Jupiters and low-mass eclipsing companions \citep{2024MNRAS.534.3893V}. To fully satisfy requirements demanding a homogeneous metallicity scale, future efforts will focus on analysing spectra for the PBS obtained either from data archives or from new observations with a single pipeline, such as the Stellar Abundances and atmospheric Parameters Pipeline \citep[SAPP,][]{gent22}, which is also used for the prime mission targets. Overall, rather than acting as a static technical checklist, the PBS catalogue provides a robust starting point that will evolve through ongoing dialogue with \plato\ work package leaders to guide future observations and resolve model degeneracies.

\subsection{Discussion}
\label{sec:discussion}

\begin{figure*}
\centering
    \includegraphics[width=\linewidth]{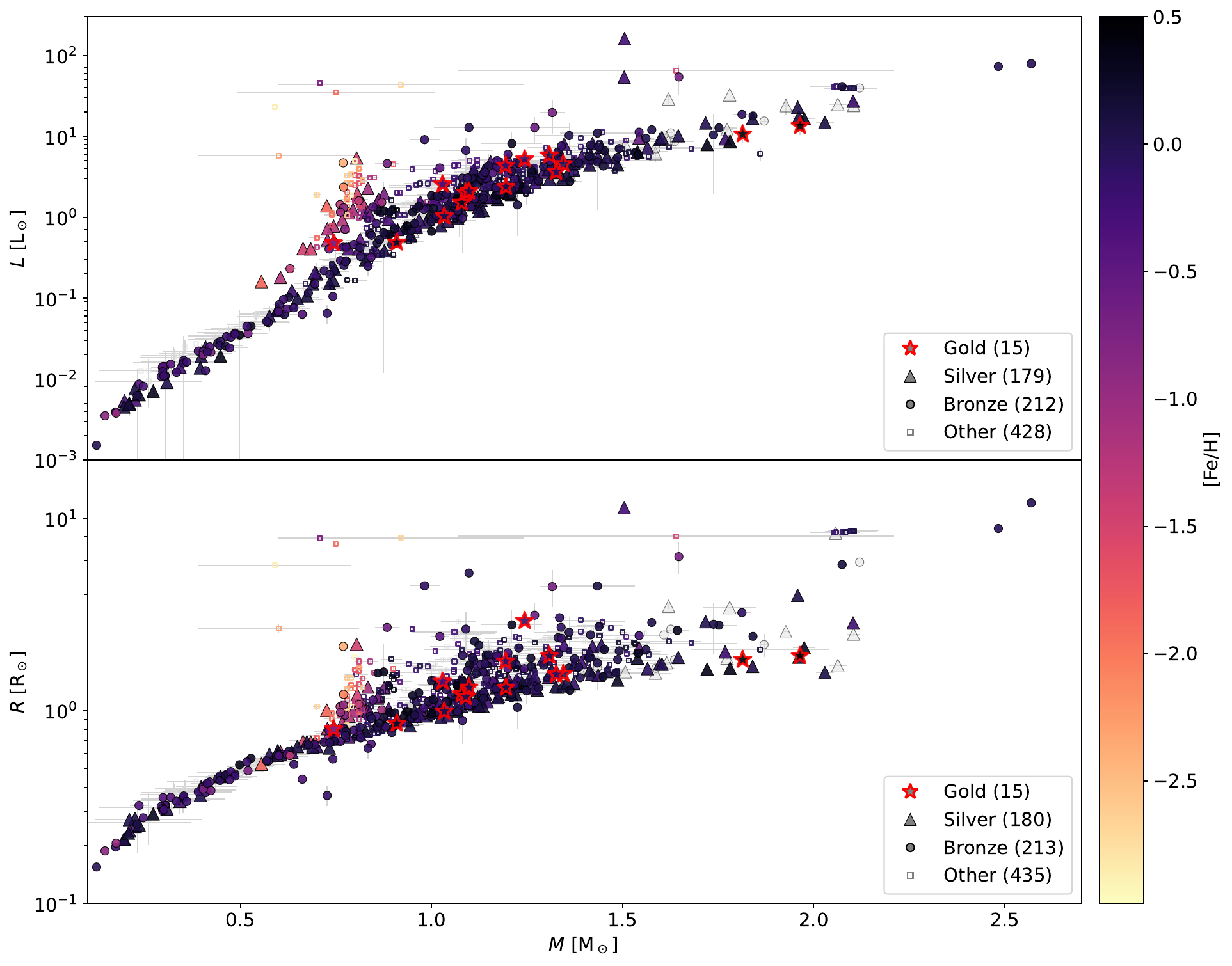}
    \caption{Mass-luminosity (top) and mass-radius (bottom) relations. Symbols as in Fig.~\ref{fig:hrd_kiel}.}
    \label{fig:mlr_mrr}
\end{figure*}

To highlight the highest-precision benchmark stars, we categorise the PBS catalogue into three quality samples: a \emph{gold} sample (15 stars with B1 mass, radius, and luminosity), a \emph{silver} sample (195 stars with B1 mass and radius), and a \emph{bronze} sample (416 stars with at least one B1 property). The \emph{gold} sample consists exclusively of binary star components, with detailed properties provided in Table~\ref{tab:gold_stars_details} and ancillary parameters available in the online catalogue. 
We emphasize that the gold sample is included in the silver sample which is included in the bronze sample.
These samples are visualised in the combined HR and Kiel diagrams (Fig.~\ref{fig:hrd_kiel}), where different symbols distinguish the benchmark quality. Figure~\ref{fig:mlr_mrr} presents the mass-luminosity and mass-radius relations. The distributions clearly outline the main-sequence stellar population, characterised by a steep increase of both luminosity and radius with stellar mass. Particularly noteworthy is the distribution of the high-precision benchmark samples: the \emph{gold} sample populates a tight, low-scatter sequence between approximately 0.7~\Msun\ and 2.0~\Msun. The stars in the M dwarfs region show very little dispersion, in particular in Fig.~\ref{fig:mlr_mrr}, and all of them belong to the \emph{silver} and \emph{bronze} samples. The expected metallicity effect is also clearly visible, making metal-poor subgiants between 0.5 and 1~\Msun\ slightly more luminous or inflated, compared to their solar-metallicity counterparts. 

\begin{table}
\centering
\footnotesize
\caption{Summary of assessed parameters for the 1090 PBS.}
\label{tab:parameter_summary}
\begin{tabular}{lccccc}
\hline\hline
Parameter & B1 & B2 & B3 & Total & N/A \\
\hline
Mass ($M$) & 249 & 500 & 91 & 840 & 250 \\
Radius ($R$) & 332 & 464 & 0 & 796 & 294 \\
Luminosity ($L$) & 168 & 575 & 0 & 743 & 347 \\
Age ($\tau$) & 0 & 452 & 27 & 479 & 611 \\
Effective Temperature (\teff) & 150 & 681 & 0 & 831 & 259 \\
Surface gravity (\logg) & 195 & 740 & 47 & 982 & 108 \\
Mean density ($\log\langle\rho\rangle$) & 204 & 538 & 45 & 787 & 303 \\
LD coefficients ($h_1', h_2'$) & 62 & 0 & 0 & 62 & 1028 \\
\hline
\multicolumn{6}{l}{\textit{Note:} Total counts include only benchmark-assessed parameters}\\
\multicolumn{6}{l}{(B1, B2, B3), excluding compiled numerical values without}\\ \multicolumn{6}{l}{ an assigned benchmark level (cf. Fig.~\ref{app:histo}).} \\
\end{tabular}
\end{table}

The metal-poor star with a mass of about 0.6~\Msun\ and \teff\ about 5000~K in the silver sample is the secondary component in a double-lined eclipsing binary system (Cl* NGC 6809 SAW V54 B). The effective temperature of this star given in \citet{2014AcA....64...11K}, based on both light-curve fitting and a theoretical colour-\teff\ calibration, seems too high for its mass and radius (see also Sect.~\ref{sec:caveats}). We note that no benchmark level is assigned to the effective temperature of this star, following the reasoning in Sect.~\ref{sec:binary}.

The limit we have imposed on the oblateness of the stars in  eclipsing binary stars for selection in the database ($\epsilon < 0.0025$) is arbitrary. It was chosen so as to have a sample size large enough to be useful while avoiding stars that are clearly unsuitable for testing stellar evolution models for single stars. 
There is no generally agreed criterion for making this decision, but we note that our criterion is more conservative than those used in other studies to select suitable benchmark stars \citep{2025ApJ...985..255D}.
More work is needed to better understand how any discrepancies between stellar models and the observed properties of these benchmark stars might be related to their binary properties. 
One encouraging recent result in the observations that AI~Phe, an eclipsing binary with an orbital  period of 24.6~days, show evidence for atomic diffusion in its photospheric abundance patterns consistent with expectations for single stars
 \citep{2026MNRAS.551g1398M}.

When using the PBS catalogue to validate theoretical frameworks, any discrepancies between model predictions and the tabulated benchmark values should be interpreted in the context of the assigned benchmark levels. A discrepant B1 property -- which is derived from direct, highly precise, and model-independent measurements -- typically points to systematic errors in the user's observational data or analytical methods rather than the benchmark value itself, though historical data quality issues in eclipsing binaries remain a known factor \citep{2010A&ARv..18...67T}. Conversely, an inaccurate B2 property, which inherently relies on stellar models with well-characterised systematic uncertainties, suggests that the underlying theoretical assumptions or input physics of the model under test require refinement. Finally, anomalous B3 properties, which are derived from broad empirical relations, most likely reflect physical peculiarities or unique evolutionary states of the individual star that cause it to deviate from standard phenomenological calibrations.

\begin{table}
\centering
\setlength{\tabcolsep}{3pt}
\caption{Crossmatch of PBS with \plato\ fields of view.}
\label{tab:plato_fov_crossmatch}
\begin{tabular}{l c c @{\,}c@{\,} c @{\,}c@{\,} c c}
\hline\hline
Field & Total & \textit{gold} & $\subset$ & \textit{silver} & $\subset$ & \textit{bronze} & Other \\
\hline
LOPN1 & 285 (26\%) & 0 & & 46 & & 111 & 174 \\
LOPS2 & 143 (13\%) & 3 & & 15 & & 27 & 116 \\
\hline
Total & 428 (39\%) & 3 & & 61 & & 138 & 290 \\
\hline
\multicolumn{8}{l}{\textit{Note:} \textit{gold}: $M, R, L = \text{B1}$; \textit{silver}: $M, R = \text{B1}$; \textit{bronze}: $\ge 1\ \text{B1}$.} \\
\multicolumn{8}{l}{ The total count is \textit{bronze} + Other.}\\
\end{tabular}
\end{table}

\subsection{Caveats}
\label{sec:caveats}

We present several caveats and limitations concerning the current database, beginning with minor issues related to data representation. For unresolved binary or multiple star systems, the $G$ and $V$ magnitudes provided in the database refer to the total apparent flux of the \gaia\ or \hipparcos\ source, as we have not computed the magnitudes of the individual components. Furthermore, we have not attempted to homogenise metallicity estimates onto a uniform solar abundance scale, and no distinction is made between literature values quoted as $\mathrm{[Fe/H]}$ or $\mathrm{[M/H]}$. This simplification is justified as most targets have solar metallicities where both scales coincide. But this represents a key limitation when using the catalogue for stellar model validation and cross-survey comparisons. In particular, for metal-poor, $\alpha$-enhanced stars, the resulting discrepancy in the bulk metallicity $Z$ reaches up to $\sim 0.3$~dex \citep{Salaris1993}. This is comparable to or exceeds typical observational uncertainties and can systematically bias theoretical stellar models that rely on precise initial compositions. Further developments to resolve these systematic scale offsets are discussed in the next section.

Regarding astrometric parameters, the zero-point correction from \citet{lindegren2021} has been applied to the \gaia\ DR3 parallaxes. However, because the calibration is limited to $G \ge 6$, the correction computed for $G = 6$ was used for any sources brighter than this limit. This will have very little impact since the corrections are small relative to the large parallaxes of nearby stars. Similarly, for 6-parameter astrometric solutions with pseudocolours outside the expected range of 1.24 to 1.72 magnitudes, the maximum corrections reached at these boundaries were applied.  Binary systems are subject to additional noise in their parallax measurements due to the orbital motion of the photocentre \citep{2021ApJ...907L..33S}. The release of the individual astrometric measurements in Gaia~DR4 will make it possible to model the astrometry in combination with the spectroscopic orbit and light curve to obtain parallaxes that exploit the full precision and accuracy of the Gaia mission.

For M dwarfs, the mass-luminosity relations used to assign B3 masses \citep[e.g.][]{mann2019} are calibrated on nearby, largely solar-metallicity systems, with calibrations rarely extending below $[\mathrm{Fe/H}] \approx -1.0$ \citep[e.g.][]{2023RAA....23e5008Q}. However, this limitation is mitigated by the very weak dependence of M-dwarf near-infrared (NIR) magnitudes on intrinsic metallicity. Indeed, stellar models confirm that the mass--NIR magnitude relation is practically independent of metallicity \citep{mann2019}. A potentially more significant, though observationally untestable, source of systematic uncertainty in these relations lies in variations of the initial helium abundance and the helium-enrichment ratio.

Calculations involving stellar rotation also carry underlying assumptions, as rotation periods are often estimated from the projected rotational velocity ($v\sin i$) or by assuming pseudo-synchronous rotation. Estimating the stellar radius from the orbital period and $v\sin i$ explicitly assumes synchronous rotation, and the derived radius will be inaccurate if this assumption fails. This method also relies on highly accurate $v\sin i$ measurements. However, $v\sin i$ measurements are subject to systematic errors of 1 to 2\,\kms\ for slow rotators due to the uncertainty in macroturbulent broadening.

Moving to detached eclipsing binary stars, \teff\ and metallicity measurements come from a wide variety of different analytical techniques. Consequently, they are generally subject to systematic errors that are difficult to quantify, which can be larger than 200\,K for \teff\ in the worst cases. The exception to this rule are direct \teff\ measurements that have been assigned a B1 benchmark level, where the systematic error is small ($\sim$10~K) and is already accounted for in the quoted standard error.

While the authors have made an effort to retain reliable high quality measurements in this work, direct interferometric measurements are not immune to systematic errors that could affect the derived angular diameters and temperatures.
A notable example is $\mu$~Cas~A, where a re-reduction of archival interferometric data yielded an angular diameter that was 8\% smaller, resulting in an effective temperature 4\% larger than previous estimates \citep{bond2020}. Consequently, one must consider possible underlying systematic errors in \teff\ from interferometry.  In one study \citep{2022ApJ...927...31T} the authors estimated it to be around 2.4\% for solar-type planet host stars, which tend to be small and faint, however more recent results (Nowacki et al., in prep.) point towards the ability to reach $2\%$ accuracy of similar targets using a fidelity criterion on the quality of the data, multi-epoch, and polychromatic observations. They also show that systematics could be controlled to a level lower than $1\%$.

The most fundamental limitation across all benchmark targets arises from the absolute flux scale. There is a minimum uncertainty of approximately 1\% in the flux scale for astronomical sources across the near-UV to near-IR wavelengths, which is where late-type stars emit the bulk of their flux. Even if we assume an absolute minimum systematic uncertainty of 0.5\% in the absolute flux scale, it dictates a minimum uncertainty in \teff\ of 0.125\% (or 7.5\,K for a 6000\,K star), while older data prior to 2020 may carry additional offsets of up to 0.87\% from the current CALSPEC flux scale \citep{2020AJ....160...21B,2025AJ....169...40B}. Despite this, some published \teff\ and bolometric flux measurements quote unrealistically small uncertainties, sometimes as low as $\pm 1$\,K. While we have chosen to quote the original published values and uncertainties in the database, we impose a strict minimum uncertainty of 1\% on $f_{\mathrm{bol}}$ whenever it is used for subsequent calculations. This fundamentally propagates to a minimum uncertainty of 0.25\% in \teff\ (equivalent to a systematic uncertainty of $\sim$15~K for a 6000\,K star) when computed from angular diameters and bolometric fluxes. These adopted floors are strongly supported by empirical validation studies comparing independent stellar radius and luminosity scales, which find similar baseline systematic limits \citep[e.g.][]{stassun2017}.

Finally, it should be understood that if a star makes it to the catalogue, this does not imply that all properties are available, or have been studied in detail. The requirement is that we have at least one robust property in order for the star to be included. Additionally, some values of $\log g$ from spectroscopy do not have an assigned benchmark level because there was no straightforward way for us to assess the accuracy of the published estimates. When these unassessed $\log g$ values are used to compute other properties (such as the stellar radius via Eq.~\ref{eqn:radfromlogg}), the resulting derived properties likewise inherit an unassigned benchmark level.

\subsection{Improvements and perspectives}
\label{sec:improvements}
The current catalogue includes data published up to May 2026; data from works published after this cutoff date are not incorporated in this release. During the work on establishing the PBS new observations along with new analyses resulted in several publications, for example  \citet{2025MNRAS.539.3741M,2025MNRAS.544.4611M}, and these results made it into the current version of the catalogue.  However, a number of other analyses are still ongoing, producing further new results, which will be reflected in an updated catalogue.  For example, for several stars, asteroseismic analyses have been conducted prior to the availability of high precision parallaxes, and re-analysing these stars, while also using more recent spectroscopic data is needed (Kaushik et al., in prep.).   Other examples include obtaining homogenous bolometric flux estimations across the full sample or relevant subsamples of data (Popa et al. in prep.). Other candidate benchmark stars have also been identified that are not in the current catalogue because observations (and/or analyses) are currently in progress. Further updates will be made once the \plato\ data will be available.  Below we describe some more specific improvements.

A major future priority is to elevate metallicity from an ancillary parameter to a formal benchmark property with defined quality levels. Due to the lack of model-independent (B1) methods, we could propose a two-tier benchmark classification: a B2 level for direct, high-resolution spectroscopic measurements, and a B3 level for estimates derived from photometric relations (particularly relevant for M dwarfs). The current version of the PBS database contains around 900 metallicity values. A transition of metallicity to a benchmark property will require to set all the metallicities on the same scale. Possible approaches are to apply corrections to the literature values based on the solar reference used, or to conduct a homogeneous spectroscopic analysis, following the strategy developed for the GBS \citep{2026A&A...705A.167C}, see also Sect.~\ref{sec:fullfilment_requirements}. Either way will involve a considerable effort from a dedicated team of experts.
 
Recent progress has been made in visible interferometry at long baselines permitting to reach a much larger sample of dwarf stars for detailed analysis of their fundamental parameters \citep{mourard2026}. A new method for the determination of the stellar radius, effective temperature, and surface gravity, combining spectroscopy, photometry, and interferometry, and based on intensity profiles from MARCS stellar atmosphere models \citep{2024A&A...691A.207E}, has been developed. First results on a small sample of stars have been obtained \citep{nayeem2026} and a publication is in preparation with a sample of $\sim$40 stars for which fundamental parameters have been determined using interferometric polychromatic data, combined with spectroscopy and stellar atmosphere models. A catalogue of a few hundreds of stars is in preparation.

Current interferometric surveys will also pave the way for determining the angular diameters of fainter stars -- those beyond the reach of direct interferometry -- by refining the SBCRs across various regions of the HR diagram. The SBCR is a powerful empirical tool that allows the derivation of a star’s angular diameter from its magnitudes in two distinct spectral bands. These relations are now widely used in multiple astrophysical contexts.  For instance, they play a key role in the characterization of exoplanet host stars \citep{dimauro22} and in asteroseismology \citep{valle24,campante19}. They have also enabled unprecedentedly precise distance measurements to eclipsing binaries in the Large \citep{pietrzynski13, pietrzynski19} and Small Magellanic Clouds \citep{graczyk20}, achieving accuracies of 1\% and 2\%, respectively. Such results are of particular importance for refining the Hubble–Lemaître constant \citep{riess22}.  In the context of the \plato\ space mission, the SBCR in (V, V$-$K) colours, initially derived by \citet{salsi2020} and later improved by \citet{salsi21}, was first integrated into the \plato\ stellar analysis pipeline \citep{gent22}. However, it was subsequently decided to adopt \gaia\ bands, which provide significantly more precise magnitudes. This led to the use also of the recent SBCR developed by \citet{kiman24}. Looking ahead, the goal is to calibrate a dedicated SBCR for \plato\ using homogeneous, multi-chromatic datasets from the CHARA interferometer \citep{mourard2026}. Preliminary work has already begun, with initial results presented in \citet{ibanez25} and \citet{ibanez26}. Ultimately, these calibrated SBCR will be used to derive the angular diameters for a future version of the \plato\ benchmark catalogue. For M dwarfs specifically, the same \citet{kiman24} sample already adopted for the \plato\ SBCR pipeline also demonstrated a link between magnetic activity and radius inflation in single M dwarfs; extending the planned CHARA-based SBCR calibration \citep{mourard2026} to include metal-poor and low-activity M-dwarf targets would help disentangle this effect from the metallicity bias affecting the wide-binary calibration sample (Sect.~\ref{sec:binary}), and move B3 mass and radius estimates for these cool dwarf stars toward B1.

It is also possible, in principle, to derive the values of ($h^{\prime}_1,h^{\prime}_2$) using long-baseline interferometry. To our knowledge, the only stars with data of sufficient quality to put useful constraints on the limb-darkening properties of F5-K7 dwarf or subgiant stars are $\alpha$~Cen~A and $\alpha$~Cen~B. \citet{kervella2017} used measurements of the fringe visibilities out to the second or third lobe in their visibility function obtained with the VLTI/PIONIER interferometer in the H band to measure the angular diameters of both stars, and studied the sensitivity of their results to the assumed form of the limb-darkening law. There is some model dependence in these results on the assumed form of the limb-darkening law, so these measurements have a benchmark level of B2. However, these measurements are not directly relevant to the benchmarking of models for the \plato\ mission because they were observed in the near-infrared H band and so they are not included in our benchmark database. However, first measurements of limb darkening in the visible domain are now reachable with the SPICA instrument (see e.g. \citet{nayeem2026, bailleul2026}). 

Concerning M~dwarfs, the current PBS sample of stars in wide binary systems with FGK stars may be extended in future versions by selecting additional stars from the literature (e.g. \citealt{montes2018}), for which dedicated spectroscopic observations are currently being carried out to provide new, homogeneous metallicity and $\alpha$-element abundance determinations (Rains et al. in prep.). Additionally, we plan to incorporate a sample of $\sim$100 M dwarfs ($11 \le G \le 15$) in wide binaries contained in the scvPIC (Sect.~3.1.3 in \citealt{zwintz2026} and \citealt{PLATO-ULG-PSM-TN-0089}).

Furthermore, wide binaries containing a main-sequence \plato\ target and a white dwarf companion offer a key testbed to validate asteroseismic ages of older field stars \citep{rebassa2021, zwintz2026}. We plan to determine the white dwarfs' cooling ages using ground-based spectroscopy and photometry (\teff\ and \logg). These independent ages will calibrate the primaries' seismic parameters and optimise theoretical model physics.

Further planned future additions to the database include results from an asteroseismic analysis of $\beta$~Aql based on SONG observations \citep{Kjeldsen2025}, the dynamical mass of $\mu^1$~Her~Aa from \citet{Marcussen2026}, asteroseismic parameters, dynamical mass, and interferometric radius for $\chi$~Dra~A \citep{2026A&A...715A..78R}, and asteroseismic parameters and dynamical mass of KIC~9693187 \citep{Beck2026}.

\section{Conclusions}
The initial \plato\ Benchmark Stars (PBS, v1.0) catalogue compiles 1090 stars spanning a wide magnitude range from $V = -1.5$ to $G = 19.8$~mag. To support pipeline validation and calibration, these targets are categorised into three high-precision quality tiers: a \emph{gold} sample of 15 stars possessing B1-level mass, radius, and luminosity; a \emph{silver} sample of 195 stars with B1 mass and radius; and a \emph{bronze} sample of 416 stars with at least one B1-level property. 
The remaining targets lack model-independent B1 parameters but are included because they feature well-characterised model-dependent (B2) or empirical (B3) properties, significantly expanding the parameter space in temperature, gravity, and metallicity available for model testing. The celestial distribution of the entire sample is well-tailored for the mission's footprints, including 285 benchmark stars in the Northern \plato\ field (LOPN1) and 143 stars within the Southern \plato\ field (LOPS2). A web interface with a TAP service is available at \url{https://astro.ulb.ac.be/pbs}.

\begin{acknowledgements}
We thank C. Loup and A. Oberto from Simbad/CDS for promptly updating the Simbad database to include any missing entries for the \plato\ benchmark stars. 
This work presents results from the European Space Agency (ESA) space mission \plato. The \plato\ payload, the \plato\ Ground Segment and \plato\ data processing are joint developments of ESA and the \plato\ Mission Consortium (PMC). Funding for the PMC is provided at national levels, in particular by countries participating in the \plato\ Multilateral Agreement (Austria, Belgium, Czech Republic, Denmark, France, Germany, Italy, Netherlands, Portugal, Spain, Sweden, Switzerland, Norway, and United Kingdom) and institutions from Brazil. Members of the \plato\ Consortium can be found at \url{https://platomission.com/}. The ESA \plato\ mission website is \url{https://www.cosmos.esa.int/plato}. We thank the teams working for \plato\ for all their work.

T.Me. is granted by the BELSPO Belgian federal research program FED-tWIN under the research profile Prf-2020-033\_BISTRO.

P.M. acknowledges support from UK Science and Technology Facilities Council (STFC) research grant numbers ST/R000638/1, ST/Y002563/1 and UKRI1193, and UK Space Agency (UKSA) grant number UKRI966.
U.H., N.M., and T.O. acknowledge support from the Swedish National Space Agency (SNSA/Rymdstyrelsen). MNL acknowledges support from the ESA PRODEX programme (PEA 4000142995). K.G.H. acknowledges support provided by the Polish National Science Center through grant no. 2023/49/B/ST9/01671.
M.B. acknowledges the Klaus Tschira Foundation.
T.Mo. acknowledges financial support from Belspo for contract PRODEX PLATO mission development.
A.R.M. acknowledges support by the Spanish MICIU grant PID2023-148661NB-I00 and by the AGAUR/Generalitat de Catalunya grant SGR-386/2021.
G.N.P. acknowledges support from the Polish National Science Center through grant 2024/53/N/ST9/03885.
J.S. acknowledges support from STFC under grant number ST/Y002563/1.
B.R-A acknowledges funding support from the ANID Basal project FB210003.
R.R. acknowledges support from Grant RYC2021-030837-I, funded by MCIN/AEI/ 10.13039/501100011033 and by “European Union NextGeneration EU/PRTR”.
This research has made use of the \simbad\ database, operated at CDS, Strasbourg, France \citep{2000A&AS..143....9W}.
This work has made use of data from the European Space Agency (ESA) mission \gaia\ (\url{https://www.cosmos.esa.int/gaia}), processed by the \gaia\ Data Processing and Analysis Consortium (DPAC, \url{https://www.cosmos.esa.int/web/gaia/dpac/consortium}).
Based on data obtained from the ESO Science Archive Facility with DOI \url{https://doi.org/10.18727/archive/33}.
This research has made use of the NASA Exoplanet Archive, which is operated by the California Institute of Technology, under contract with the National Aeronautics and Space Administration under the Exoplanet Exploration Program.
This research has made use of data obtained from the 4XMM XMM-Newton serendipitous source catalogue compiled by the XMM-Newton Survey Science Centre consortium.

\end{acknowledgements}

\bibliographystyle{aa}
\bibliography{aa63390-26}

\appendix

\section{Stellar parameters relations and uncertainties}
\label{app:params}
\begin{itemize}
\item Stellar radius $R$ from angular diameter $\diam$ and from parallax $\varpi$: 
\begin{equation}
\label{eqn:radfromang}
\frac{R}{R_\odot} = \frac{1}{2}\frac{\diam}{\varpi}\frac{\mathrm{au}}{R_\odot}
\end{equation}
where $\diam$ and $\varpi$ are in units of mas, and the solar radius $R_\odot$ and the astronomical unit, au, are given in Table~\ref{tab:ref}. The associated relative uncertainty is given by:
\begin{equation}
\frac{\Delta R}{R} = \sqrt{\left(\frac{\Delta \diam}{\diam}\right)^2 + \left(\frac{\Delta \varpi}{\varpi}\right)^2}
\end{equation}
\item Stellar radius $R$ from stellar mass $M$ and \logg: 
\begin{equation}
\label{eqn:radfromlogg}
\frac{R}{R_\odot} = 10^{-\frac{1}{2}(\log g - \log g_\odot)} \sqrt{\frac{M}{M_\odot}}
\end{equation}
where the solar radius $R_\odot$, surface gravity $\log g_\odot$ and mass \Msun\ are given in Table~\ref{tab:ref}. The associated relative uncertainty is given by:
\begin{equation}
\frac{\Delta R}{R} = \frac{1}{2} \sqrt{\left(\frac{\Delta M}{M}\right)^2 + (\ln 10\ \Delta \log g)^2}
\end{equation}
\item Angular diameter $\diam$ from stellar radius $R$ and parallax $\varpi$: 
\begin{equation}
\label{eqn:angfromrad}
\diam = 2\varpi R\frac{R_\odot}{\mathrm{au}}
\end{equation}
where $\varpi$ are in units of mas, $R$ in solar units, and $R_\odot$ and the astronomical unit, au, are given in Table~\ref{tab:ref}. The associated relative uncertainty is given by:
\begin{equation}
\frac{\Delta \diam}{\diam} = \sqrt{\left(\frac{\Delta R}{R}\right)^2 + \left(\frac{\Delta \varpi}{\varpi}\right)^2}
\end{equation}
\item Surface gravity \logg\ from stellar mass $M$ and radius $R$:
\begin{equation}
\label{eqn:loggfrommasandrad}
\log g = \log g_\odot + \log{\frac{M}{M_\odot}} - 2 \log\frac{R}{R_\odot}
\end{equation}
where the solar surface gravity \logg$_\odot$, mass \Msun\ and radius $R_\odot$ are given in Table~\ref{tab:ref}. The associated absolute uncertainty in \logg\ (corresponding to the relative uncertainty in $g$) is given by:
\begin{equation}
\Delta \log g = \frac{1}{\ln 10} \sqrt{\left(\frac{\Delta M}{M}\right)^2+ \left(2\frac{\Delta R}{R}\right)^2}
\end{equation}
\item Stellar mass $M$ from surface gravity \logg\ and radius $R$:
\begin{equation}
\frac{M}{M_\odot} = 10^{\left(\log g -\log g_\odot\right)} \left(\frac{R}{R_\odot}\right)^2
\end{equation}
where the solar mass \Msun, surface gravity \logg$_\odot$, and radius $R_\odot$ are given in Table~\ref{tab:ref}. The associated relative uncertainty is given by:
\begin{equation}
\frac{\Delta M}{M} = \sqrt{\left(\ln 10 \Delta \log g \right)^2+ \left(2\frac{\Delta R}{R}\right)^2}
\end{equation}
\item The logarithm of the mean stellar density \rhostar\ from stellar mass $M$ and radius $R$:
\begin{equation}
\log \left(\frac{\rhostar}{\rhosun}\right) = \log\left(\frac{M}{M_\odot}\right) - 3 \log\left(\frac{R}{R_\odot}\right)
\end{equation}
where the solar mean density \rhosun, mass \Msun\ and radius $R_\odot$ are given in Table~\ref{tab:ref}. The associated absolute uncertainty in the logarithm (relative uncertainty in \rhostar) is given by:
\begin{equation}
\Delta \log\left(\frac{\rhostar}{\rhosun}\right) = \frac{1}{\ln 10}\sqrt{\left( \frac{\Delta M}{M}\right)^2 + \left( 3\frac{\Delta R}{R}\right)^2}
\end{equation}
\item Luminosity $L$ from bolometric flux $f_\mathrm{bol}$ and from parallax $\varpi$: 
\begin{equation}
\label{eqn:lumfromfbol}
\frac{L}{L_\odot} = \frac{1.679616 \times 10^{7}}{\pi} \left(\frac{\mathrm{au}}{\varpi }\right)^2  \frac{f_\mathrm{bol}}{L_\odot}
\end{equation}
where $f_\mathrm{bol}$ is expressed in $10^{-11}$~W~m$^{-2}$ or equivalently in $10^{-8}$ erg~s$^{-1}$~cm$^{-2}$ and $\varpi$ in mas. The solar luminosity $L_\odot$, and the astronomical unit, au, are given in Table~\ref{tab:ref}. The associated relative uncertainty is given by:
\begin{equation}
\frac{\Delta L}{L} = \sqrt{\left(2 \frac{\Delta \varpi}{\varpi}\right)^2 + \left(\frac{\Delta f_\mathrm{bol}}{f_\mathrm{bol}}\right)^2}
\end{equation}
\item Luminosity $L$ from effective temperature \teff\ and radius $R$: 
\begin{equation}
\label{eqn:lumfromteff}
    \frac{L}{L_\odot} = \left(\frac{R}{R_\odot}\right)^2 \left(\frac{T_\mathrm{eff}}{T_\mathrm{eff}^\odot}\right)^4
\end{equation}
where the solar luminosity $L_\odot$, radius $R_\odot$ and effective temperature \teff$^\odot$ are given in Table~\ref{tab:ref}. The associated relative uncertainty is given by:
\begin{equation}
\frac{\Delta L}{L} = \sqrt{\left(2\frac{\Delta R}{R}\right)^2 + \left(4\frac{\Delta T_\mathrm{eff}}{T_\mathrm{eff}}\right)^2}
\end{equation}
\item Bolometric flux $f_\mathrm{bol}$ from luminosity $L$ and parallax $\varpi$:
\begin{equation}
\label{eqn:fbolfromlum}
f_\mathrm{bol} = \frac{\pi}{1.679616 \times 10^{7}}\left(\frac{\varpi}{\mathrm{au}}\right)^2 \left(\frac{L}{L_\odot}\right){L_\odot}
\end{equation}
where $f_\mathrm{bol}$ is expressed in $10^{-11}$~W~m$^{-2}$ or equivalently in $10^{-8}$ erg~s$^{-1}$~cm$^{-2}$ and $\varpi$ in mas. The solar luminosity $L_\odot$, and the astronomical unit, au, are given in Table~\ref{tab:ref}. The associated relative uncertainty is given by:
\begin{equation}
\frac{\Delta f_\mathrm{bol}}{f_\mathrm{bol}} = \sqrt{\left(\frac{\Delta L}{L}\right)^2 + \left(2 \frac{\Delta \varpi}{\varpi}\right)^2}
\end{equation}
\item Effective temperature \teff\ from angular diameter $\diam$ and bolometric flux $f_\mathrm{bol}$: 
\begin{equation}
\label{eqn:tefffromangandfbol}
T_\mathrm{eff} = \left(\frac{10^{-11}f_\mathrm{bol}}{\sigma_{\rm SB}}\right)^{1/4} \left(\frac{\diam}{2\cdot 3.6\times10^6}\frac{\pi}{180}\right)^{-1/2} 
\end{equation}
where $T_\mathrm{eff}$ is in K, $f_\mathrm{bol}$ is expressed in $10^{-11}$~W~m$^{-2}$ or equivalently in $10^{-8}$ erg~s$^{-1}$~cm$^{-2}$ and $\diam$ in mas, and the Stefan-Boltzmann constant $\sigma_{\rm SB}$ is given in Table~\ref{tab:ref}. The associated relative uncertainty is given by:
\begin{equation} 
\frac{\Delta T_\mathrm{eff}}{T_\mathrm{eff}} = \sqrt{\left(\frac{1}{4}\frac{\Delta f_\mathrm{bol}}{f_\mathrm{bol}} \right)^2 +  \left(\frac{1}{2}\frac{\Delta \diam}{\diam} \right)^2}
\end{equation}
\item Effective temperature \teff\ from luminosity $L$ and radius $R$:
\begin{equation}
T_\mathrm{eff} = \left(\frac{R_\odot}{R}\right)^{1/2} \left(\frac{L}{L_\odot} \right)^{1/4} T_\mathrm{eff}^\odot
\end{equation}
where the solar radius $R_\odot$, luminosity $L_\odot$, and effective temperature \teff$^\odot$ are given in Table~\ref{tab:ref}. The associated relative uncertainty is given by:
\begin{equation} 
\frac{\Delta T_\mathrm{eff}}{T_\mathrm{eff}} =  \sqrt{\left(\frac{1}{2}\frac{\Delta R}{R} \right)^2 +  \left(\frac{1}{4}\frac{\Delta L}{L} \right)^2}
\end{equation}
\item Radius from logarithm of the mean stellar density \rhostar\  and mass, $M$: 
\begin{equation}
\log\left(\frac{R}{R_\odot}\right) = \frac{1}{3}\log\left(\frac{M}{M_\odot}\right) -\frac{1}{3}\log \left(\frac{\langle\rho\rangle}{\langle\rho\rangle_\odot}\right)  
\end{equation}
where the solar mean density $\bar\rho_\odot$, mass \Msun\ and radius $R_\odot$ are given in Table~\ref{tab:ref}. The associated relative uncertainty is given by:
\begin{equation}
\frac{\Delta R}{R} = \sqrt{\left( \frac{\Delta M}{3M}\right)^2 + \left(\frac{\ln10\,\Delta\log(\rhostar/\rhostar_{\odot})}{3} \right)^2}
\end{equation}
\end{itemize}

\section{Reference constant values}
\begin{table*}
\caption{Reference constant values.}
\label{tab:ref}
\centering
\begin{tabular}{lcrlr}
\hline \hline
Physical property & Symbol & Value & Unit & Reference \\
\hline
Solar mass parameter & $G$\Msun & 1.3271244$\times10^{20}$ & m$^{3}$s$^{-2}$ & (1) \\
Solar mass & \Msun & 1.98841$\times10^{30}$ & kg & (2) \\
Solar radius & $R_\odot$ & 6.957$\times10^8 $  & m & (1) \\
Solar luminosity & $L_\odot$ & 3.828$\times10^{26}$ & W & (1) \\
Solar effective temperature & \teff$^\odot$ & 5772 & K & (1) \\
Solar surface gravity & \logg$_\odot$ & 4.4381 & $\log{\left(\mathrm{cm~s}^{-2}\right)}$ & (3) \\
Solar mean density & \rhosun\ & 1410 & kg m$^{-3}$ & (4) \\
Newtonian constant of gravitation & $G$ & 6.67430$\times10^{-11}$ & m$^{3}$kg$^{-1}$s$^{-2}$ & (5) \\
Astronomical unit of length &   au & 1.49597870700$\times10^{11}$ & m & (6) \\
Parsec & pc & 3.08567758147$\times10^{16}$ & m & (7) \\
Stefan-Boltzmann constant & $\sigma_{\rm SB}$ & 5.670374419$\times10^{-8}$ & Wm$^{-2}$K$^{-4}$ & (5) \\
Solar frequency at maximum power& $\nu_{\text{max},\odot}$& 3090 & $\mu$Hz & (8)\\
Solar large frequency separation & $\Delta\nu_\odot$ & 135.1 & $\mu$Hz & (8)\\
\hline\hline
   \end{tabular}
   \tablebib{
   (1) Resolution B3 adopted at IAU General Assembly 2015 \citep{prsa2016};
   (2) calculated from $G$\Msun\ and $G$;
   (3) calculated from $G$\Msun\ and $R_\odot$;
   (4) calculated from \Msun\ and $R_\odot$;
   (5) 2022 CODATA recommended value \citep{2025RvMP...97b5002M};
   (6) Resolution B2 adopted at IAU General Assembly 2012;
   (7) calculated from au;
   (8) from \citet{2011ApJ...743..143H}.
   }
\end{table*}

\section{Data compilations}
\begin{table*}
\caption{Summary of the observational data compilations used to construct the \plato\ benchmark stars database, organized by primary measurement method. The table lists the target types, the number of stars included ($N$), and the resulting benchmark properties with their assessed levels.}
\label{tab:data_compilation_summary}
\centering
\footnotesize
\begin{tabular}{p{0.30\linewidth} p{0.23\linewidth} c p{0.16\linewidth} p{0.17\linewidth}}
\hline\hline
Source & Stellar type & $N$ & Primary method & Properties (level) \\
\hline
\multicolumn{5}{c}{\textbf{Binary and multiple systems}} \\
\hline
DEBCat \citep{southworth2015} & Detached EBs & 159 & EB + SB2 & $M$ (B1), $R$ (B1) \\
Araucaria Project \citep{2023AA...672A.119G} & Detached EBs & 10 & Orbital parallax + SB2 & $M$ (B1) \\
EBLM Systems \newline \citep{maxted2023, 2025MNRAS.539.3741M} & EBs with low-mass secondary & 36 & Transit light curves + SB1 & $\langle\rho\rangle$ (B1), $R$ (B1/B2) \\
Review on masses  \citep{serenelli2021} & Visual and detached binaries & 110 & Dynamical masses & $M$ (B1) \\
M dwarfs in wide binaries \citep[e.g.,][]{Rains2021, Olander2025a} & M dwarf companions & 26 & Empirical relations & $M$ (B3), $R$ (B3), $T_{\text{eff}}$ (B2), [Fe/H] \\
\hline
\multicolumn{5}{c}{\textbf{Interferometry and astrometry}} \\
\hline
GBS v3  \citep{soubiran2024} & FGK main sequence \& subgiant stars & 105 & Interferometry + \newline Parallax & $T_{\text{eff}}$ (B1), $R$ (B1), $L$ (B1) \\
M-dwarf compilation  \citep[e.g.,][]{boyajian2012, Rabus2019} & M dwarfs & 39 & Interferometry & $R$ (B1), $T_{\text{eff}}$ (B1) \\
\hline
\multicolumn{5}{c}{\textbf{Asteroseismic samples}} \\
\hline
\gaia-Kepler properties  \citep{berger2020} & FGK dwarfs & 29 & Asteroseismology + Photometry & $M$ (B2), $R$ (B2), $\tau$ (B2) \\
APOGEE-Kepler  \citep{claytor2020} & Cool dwarfs & 19 & Asteroseismology + Spectroscopy & $M$ (B2), $R$ (B2), $\tau$ (B2) \\
KEYSTONE sample \citep{Lund2016,2024AA...688A..13L} & Dwarf \& subgiant oscillators &170 & Asteroseismology & $M$ (B2), $R$ (B2), $\tau$ (B2) \\
\hline
\multicolumn{5}{c}{\textbf{Large surveys}} \\
\hline
Titans I \& II  \citep{giribaldi2021, giribaldi2023} & Metal-poor dwarfs \& red giants & 55 & 3D non-LTE $H\alpha$ fitting & $T_{\text{eff}}$ (B2) \\
SPOCS catalogue  \citep{valenti2005} & Bright FGK dwarfs & 9 & Spectroscopy + Isochrones & $\tau$ (B2), $M$ (B2) \\
\gaia\ DR3 Golden Sample \newline  \citep{2023AA...674A..39G} & FGKM dwarfs and subgiants, incl. solar analogues & 131 & High-res Spectroscopy & $T_{\text{eff}}$, \logg, [Fe/H] \\
\hline
\end{tabular}
\tablefoot{
$N$ represents the number of stars strictly fulfilling the \plato\ core mission requirements that were selected from the parent catalogues. 
}
\end{table*}
\section{X-ray properties of M dwarfs}
\begin{table*}
\caption{X-ray properties of benchmark M dwarfs.}
\label{tab:mdwarfs_xray}
\centering
\begin{tabular}{lllclll}
\hline\hline
Primary ID & Alternative ID & $C_X$ & $\sigma (C_X)$ & $\log(f_X/f_{\mathrm{bol}})$ & $\sigma (\log(f_X/f_{\mathrm{bol}}))$ & Source \\
\hline
\multicolumn{7}{c}{Stars in the interferometry sample} \\
\hline
V* GX And & GJ 15A & 0.288 & 0.026 & $-4.51$ & 0.04 & 2RXS \\
HD  36395 & GJ 205 & \dots & \dots & $-4.97$ & 0.06 & ZP2025 \\
HD  79210 & GJ 338A & \dots & \dots & $-4.22$ & 0.5 & R+2022 \\
HD  79211 & GJ 338B & \dots & \dots & $-4.69$ & 0.5 & R+2022 \\
TYC 3012-2528-1 & BD+44 2051A & \dots & \dots & $-5.4$ & 0.1 & ZP2025 \\
V* FI Vir & Ross 128 & \dots & \dots & $-4.8$ & 0.1 & ZP2025 \\
HD 119850 & GJ 526 & \dots & \dots & $-5.5$ & 0.1 & ZP2025 \\
TYC 4428-1943-1 & BD+68 946 & \dots & \dots & $-4.83$ & 0.03 & ZP2025 \\
HD 173740 & GJ 725B & \dots & \dots & $-4.62$ & 0.05 & ZP2025 \\
V* V1428 Aql & HD 180617 & \dots & \dots & $-5.32$ & 0.01 & ZP2025 \\
HD 199305 & GJ 809 & \dots & \dots & $-4.84$ & 0.05 & ZP2025 \\
HD 216899 & GJ 880 & \dots & \dots & $-5.17$ & 0.5 & R+2022 \\
HD  95735 & GJ 411 & 1.9007 & 0.0091 & $-4.574$ & 0.003 & 4XMM \\
HD 285968 & GJ 176 & \dots & \dots & $-5.28$ & 0.03 & ZP2025 \\
TYC 1984-2613-1 & Ross 905 & \dots & \dots & $-5.71$ & 0.04 & ZP2025 \\
V* HO Lib & BD-07 4003 & \dots & \dots & $-5.22$ & 0.03 & ZP2025 \\
TYC 2063-479-1 & BD+25 3173 & \dots & \dots & $-3.64$ & 0.5 & R+2022 \\
V* V2500 Oph & Barnard's star & 0.0234 & 0.0026 & $-5.40$ & 0.05 & 2RXP \\
V* IL Aqr & BD-15 6290 & \dots & \dots &$ -4.83$ & 0.02 & ZP2025 \\
HD 225213 & GJ 1 & 0.0197 & 0.0024 & $-5.52$ & 0.05 & 2RXP \\
TYC  246-1068-1 & BD+01 2447 & \dots & \dots & $-5.1$ & 0.1 & ZP2025 \\
* alf Cen C & Proxima Centauri & 5.475 & 0.013 & $-3.58$ & 0.03 & 4XMM \\
TYC 8346-37-1 & CD-46 11540 & \dots & \dots & $-4.19$ & 0.01 & ZP2025 \\
HD 204961 & GJ 832 & 0.0850 & 0.0021 & $-5.46$ & 0.02 & 4XMM \\
TYC  173-3208-1 & BD+05 1668 & \dots & \dots & $-5.41$ & 0.01 & ZP2025 \\
V* V2306 Oph & BD-12 4523 & \dots & \dots & $-5.09$ & 0.5 & R+2022 \\
V* V1216 Sgr & Ross 154 & \dots & \dots & $-3.34$ & 0.01 & ZP2025 \\
HD 217987 & GJ 887 & \dots & \dots & $-4.95$ & 0.01 & ZP2025 \\
TYC  882-1111-1 & Wolf 437 & 0.00672 & 0.00071 & $-5.81$ & 0.05 & 4XMM \\
HD 265866 & GJ 251 & \dots & \dots & $-5.4$ & 0.2 & ZP2025 \\
TYC 1978-1286-1 & Ross 104 & \dots & \dots & $-5.0$ & 0.1 & ZP2025 \\
TYC 3878-1193-1 & G 202-48 & \dots & \dots & $-4.88$ & 0.04 & ZP2025 \\
HD 165222 & GJ 701 & \dots & \dots & $-5.0$ & 0.2 & ZP2025 \\
HD 209290 & GJ 846 & \dots & \dots & $-5.27$ & 0.5 & R+2022 \\
V* EV Lac & GJ 873 & 11.373 & 0.027 & $-3.01$ & 0.01 & 4XMM \\
V* EQ Peg A & BD+19 5116A & \dots & \dots & $-2.78$ & 0.5 & R+2022 \\
V* HH And & Ross 248 & \dots & \dots & $-3.75$ & 0.06 & ZP2025 \\
V* BR Psc & GJ 908 & \dots & \dots & $-5.2$ & 0.1 & ZP2025 \\
\hline
\multicolumn{7}{c}{Stars in the wide-binary sample} \\
\hline
V* BX Cet & GJ 105B & \dots & \dots & $-4.5$ & \dots & ZP2025 \\
TYC 1598-325-1 & Ross 730 & 0.002 & \dots & $-6.3$ & \dots & C+2023 \\
HD 349726 & GJ 745B & 0.015 & 0.005 & $-5.5$ & 0.1 & C+2023 \\
TYC 4959-178-1 & Ross 486A & 0.076 & 0.027 & $-4.9$ & 0.2 & eRASS1 \\
* rho01 Cnc B & GJ 324B & 0.054 & 0.029 & $-4.5$ & 0.2 & eRASS1 \\
2MASS J18244689-0620311 & BD-06 4756B & 0.083 & 0.018 & $-2.9$ & 0.1 & 2RXS \\
2MASS J20032651+2952000 & G 125-55 & 0.0614 & 0.0067 & $-3.87$ & 0.05 & 4XMM \\
HD 233153 & GJ 212 & \dots & \dots & $-3.79$ & 0.5 & R+2022 \\
HD  61606B & BD-03 2002 & 0.395 & 0.074 & $-4.68$ & 0.08 & eRASS1 \\
TYC 4682-2798-1 & G 70-50 & 0.0413 & 0.0035 & $-4.79$ & 0.06 & 4XMM \\
HD 154363B & PM J17052-0505 & \dots & \dots & $-4.82$ & 0.5 & R+2022 \\
HD  50281B & GJ 250B & \dots & \dots & $-5.25$ & 0.02 & ZP2025 \\
HD  24916B & GJ 157B & 0.4647 & 0.039 & $-2.29$ & 0.05 & 2RXS \\
\hline
\end{tabular}
\tablefoot{$C_X$ and $\sigma (C_X)$ are X-ray count rates and their uncertainties in counts s$^{-1}$ and $f_X/f_{\mathrm{bol}}$ is the X-ray to bolometric flux ratio. Values without uncertainties are upper limits. See text for details (Sect.~\ref{sec:methods-Md}).}
\tablebib{ZP2025: \citet{2025A&A...694A..93Z}, R+2022: \citet{2022A&A...662A..41R}, C+2023: \citet{2023A&A...676A..14C}, 4XMM /  2RXP / eRASS1 / 2RXS: this work, using data from XMM-Newton, ROSAT, or eROSITA.}
\end{table*}

\section{The gold sample}

\begin{table*}[htbp]
\centering
\footnotesize
\caption{Detailed physical parameters for the 15 stars in the \textit{gold} sample (having B1 levels of mass, radius, and luminosity).}
\label{tab:gold_stars_details}
\begin{tabular}{lcccc}
\hline\hline
Star & $M$ [\Msun] & $R$ [\Rsun] & $L$ [\Lsun] & $\tau$ [Gyr] \\
     & \teff\ [K] & \logg\ [dex] & $\log(\rho/\rho_{\odot})$ [dex] & LD coefficients \\
     & [Fe/H] [dex] & $P_{\text{rot}}$ [d] & $f_{\text{bol}}$ [$10^{-11} \text{ W m}^{-2}$] & $\varpi$ [mas] \\
\hline
\multirow{3}{*}{$\alpha$ Cen A} & $1.0788 \pm 0.0029$~[1] & $1.2175 \pm 0.0055$~(A.1) & $1.514 \pm 0.015$~(A.13) & $7.1 \pm 1.1$~[2] \\
 & $5803 \pm 19$~(A.19) & $4.3001 \pm 0.0041$~(A.7) & $-0.2235 \pm 0.006$~(A.11) & -- \\
 & $0.237 \pm 0.007$~[3] & $28.8 \pm 2.5$~[4] & $2730.9 \pm 5.4$~[5] & $750.81 \pm 0.38$~[1] \\
\hline
\multirow{3}{*}{$\alpha$ Cen B} & $0.9092 \pm 0.0025$~[1] & $0.8591 \pm 0.0036$~(A.1) & $0.4882 \pm 0.0049$~(A.13) & $7.1 \pm 1.1$~[2] \\
 & $5206 \pm 17$~(A.19) & $4.5287 \pm 0.0038$~(A.7) & $0.1566 \pm 0.0056$~(A.11) & -- \\
 & $0.221 \pm 0.016$~[3] & $43.3 \pm 1.5$~[6] & $880.5 \pm 3.5$~[5] & $750.81 \pm 0.38$~[1] \\
\hline
\multirow{3}{*}{EBLM J0113+31 A$^\dagger$} & $1.029 \pm 0.025$~[7] & $1.417 \pm 0.014$~[7] & $2.55 \pm 0.08$~[7] & $6.7 \pm 0.5$~[7] \\
 & $6124 \pm 50$~[7] & $4.148 \pm 0.006$~[7] & $-0.4413 \pm 0.0072$~[7] & -- \\
 & $-0.33 \pm 0.15$~[7] & $10.87 \pm 0.6$~[7] & $0.262 \pm 0.005$~[7] & $5.665 \pm 0.031$~[8] \\
\hline
\multirow{3}{*}{EBLM J0608-59 A$^\dagger$} & $1.098 \pm 0.018$~[9] & $1.321 \pm 0.017$~[9] & $2.09 \pm 0.09$~[9] & $6 \pm 0.3$~[9] \\
 & $6031 \pm 56$~[9] & $4.24 \pm 0.01$~[9] & $-0.322 \pm 0.016$~[9] & $h_1' = 0.88 \pm 0.002, h_2' = 0.16 \pm 0.02$~[10] \\
 & $0.01 \pm 0.05$~[9] & $18.6 \pm 3.1$~[11] & $0.0417 \pm 0.0005$~[9] & $2.502 \pm 0.01$~[8] \\
\hline
\multirow{3}{*}{HD 22064 A$^\dagger$} & $1.345 \pm 0.031$~[12] & $1.554 \pm 0.014$~[12] & $4.55 \pm 0.06$~[12] & $1.8 \pm 0.3$~[12] \\
 & $6763 \pm 49$~[12] & $4.184 \pm 0.006$~[12] & $-0.4449 \pm 0.0073$~[12] & -- \\
 & $-0.03 \pm 0.15$~[12] & $3.09 \pm 0.03$~[12] & $0.751 \pm 0.009$~[12] & $7.178 \pm 0.021$~[8] \\
\hline
\multirow{3}{*}{$\iota$ Peg A} & $1.326 \pm 0.016$~[13] & $1.543 \pm 0.068$~(A.1) & $3.638 \pm 0.085$~(A.13) & -- \\
 & $6418 \pm 142$~(A.19) & $4.15 \pm 0.06$~[5] & $-0.471 \pm 0.058$~(A.11) & -- \\
 & $-0.11 \pm 0.02$~[14] & -- & $82.21 \pm 0.33$~[5] & $86.9 \pm 1$~[13] \\
\hline
\multirow{3}{*}{$\mu$ Cas A} & $0.744 \pm 0.012$~[15] & $0.8027 \pm 0.0079$~(A.1) & $0.478 \pm 0.0076$~(A.13) & $12.7 \pm 2.7$~[15] \\
 & $5357 \pm 31$~(A.19) & $4.515 \pm 0.011$~[15] & $0.14 \pm 0.013$~(A.11) & -- \\
 & $-0.83 \pm 0.02$~[14] & -- & $25.98 \pm 0.38$~[5] & $132.66 \pm 0.69$~[15] \\
\hline
\multirow{3}{*}{TYC 8511-888-1 A$^\dagger$} & $1.3083 \pm 0.0051$~[16] & $1.9289 \pm 0.0031$~[16] & $5.83 \pm 0.15$~[16] & -- \\
 & $6457 \pm 45$~[16] & $3.9836 \pm 0.0012$~[16] & $-0.7392 \pm 0.0027$~(A.11) & -- \\
 & $0 \pm 0.2$~[16] & $27.13 \pm 22.65$~[17] & $0.1294 \pm 0.0036$~(A.17) & $2.633 \pm 0.014$~[8] \\
\hline
\multirow{3}{*}{TYC 8511-888-1 B$^\dagger$} & $1.0897 \pm 0.0035$~[16] & $1.1814 \pm 0.0038$~[16] & $2.018 \pm 0.056$~[16] & -- \\
 & $6324 \pm 44$~[16] & $4.3297 \pm 0.0026$~[16] & $-0.1799 \pm 0.0044$~(A.11) & -- \\
 & $0 \pm 0.2$~[16] & $28.49 \pm 13.66$~[17] & $0.0448 \pm 0.0013$~(A.17) & $2.633 \pm 0.014$~[8] \\
\hline
\multirow{3}{*}{AI Phe A} & $1.1938 \pm 0.0008$~[18] & $1.8036 \pm 0.0022$~[18] & $4.329 \pm 0.063$~[19] & $4.39 \pm 0.32$~[20] \\
 & $6199 \pm 22$~[19] & $4.002 \pm 0.0011$~[18] & $-0.6925 \pm 0.0016$~[18] & -- \\
 & $-0.14 \pm 0.1$~[19] & -- & $0.4797 \pm 0.0076$~(A.17) & $5.885 \pm 0.019$~[19] \\
\hline
\multirow{3}{*}{AI Phe B} & $1.2438 \pm 0.0008$~[18] & $2.9303 \pm 0.0023$~[18] & $5.207 \pm 0.065$~[19] & $4.39 \pm 0.32$~[20] \\
 & $5094 \pm 16$~[19] & $3.5981 \pm 0.0009$~[18] & $-1.3073 \pm 0.0012$~[18] & -- \\
 & $-0.14 \pm 0.1$~[19] & -- & $0.577 \pm 0.0081$~(A.17) & $5.885 \pm 0.019$~[19] \\
\hline
\multirow{3}{*}{LL Aqr A} & $1.1947 \pm 0.0009$~[21] & $1.318 \pm 0.0013$~[21] & $2.377 \pm 0.077$~[21] & $4.7 \pm 0.14$~[22] \\
 & $6242 \pm 50$~[21] & $4.2755 \pm 0.0009$~[21] & $-0.2825 \pm 0.0013$~(A.11) & -- \\
 & $0.02 \pm 0.04$~[23] & $19.1 \pm 2.7$~[23] & $0.409 \pm 0.013$~(A.17) & $7.358 \pm 0.02$~[24] \\
\hline
\multirow{3}{*}{LL Aqr B} & $1.0334 \pm 0.0006$~[21] & $0.9927 \pm 0.0008$~[21] & $1.03 \pm 0.031$~[21] & $4.7 \pm 0.14$~[22] \\
 & $5839 \pm 44$~[21] & $4.4587 \pm 0.0008$~[21] & $0.0238 \pm 0.0011$~(A.11) & -- \\
 & $0.03 \pm 0.06$~[23] & $14.1 \pm 1.6$~[23] & $0.1771 \pm 0.0054$~(A.17) & $7.358 \pm 0.02$~[24] \\
\hline
\multirow{3}{*}{WW Aur A$^\dagger$} & $1.964 \pm 0.007$~[25] & $1.927 \pm 0.011$~[25] & $13.4 \pm 3.1$~(A.15) & -- \\
 & $7962 \pm 452$~[25] & $4.1614 \pm 0.0052$~(A.7) & $-0.5615 \pm 0.0076$~(A.11) & -- \\
 & $0.5 \pm 0.25$~[25] & $2.79 \pm 0.26$~[25] & $5.4 \pm 1.2$~(A.17) & $11.186 \pm 0.069$~[8] \\
\hline
\multirow{3}{*}{WW Aur B$^\dagger$} & $1.814 \pm 0.007$~[25] & $1.841 \pm 0.011$~[25] & $10.6 \pm 2.4$~(A.15) & -- \\
 & $7674 \pm 436$~[25] & $4.1666 \pm 0.0055$~(A.7) & $-0.5365 \pm 0.008$~(A.11) & -- \\
 & $0.5 \pm 0.25$~[25] & $2.52 \pm 0.22$~[25] & $4.24 \pm 0.97$~(A.17) & $11.186 \pm 0.069$~[8] \\
\hline
\multicolumn{5}{l}{\textit{Note:}
For each star, the first two rows list the eight assessed stellar properties, while the third row lists selected ancillary parameters.} \\
\multicolumn{5}{l}{Other ancillary parameteters are available \href{https://astro.ulb.ac.be/pbs}{online}.}\\
\multicolumn{5}{l}{$^\dagger$ Component not resolved by \gaia~DR3 (the reported parallax corresponds to the unresolved system).} \\
\multicolumn{5}{l}{\textbf{References:}} \\
\multicolumn{5}{l}{[1] \citet{2021AJ....162...14A}; [2] \citet{2021AA...646A...7S}; [3] \citet{morel2018}; [4] \citet{1991ApJ...372..610H}; [5] \citet{soubiran2024}} \\
\multicolumn{5}{l}{[6] \citet{1993AA...276...78C}; [7] \citet{2022MNRAS.513.6042M}; [8] \citet{2022yCat.1355....0G}; [9] \citet{2024MNRAS.531.4577M}; [10] \citet{2025MNRAS.537.3943F}} \\
\multicolumn{5}{l}{[11] \citet{kostov2020}; [12] \citet{2023MNRAS.522.2683M}; [13] \citet{1999ApJ...515..356B}; [14] \citet{2022yCat..36630004S}; [15] \citet{bond2020}} \\
\multicolumn{5}{l}{[16] \citet{2022MNRAS.517.5129M}; [17] \citet{2021MNRAS.500.4972R}; [18] \citet{maxted2020}; [19] \citet{miller2020}; [20] \citet{2016AA...591A.124K}} \\
\multicolumn{5}{l}{[21] \citet{2026MNRAS.545f2078M}; [22] \citet{2023AA...678A.203V}; [23] \citet{graczyk2016}; [24] \citet{2023AA...672A.119G}; [25] \citet{2005MNRAS.363..529S}} \\
\end{tabular}
\end{table*}

\section{Distribution of benchmark stellar properties}

\begin{figure*}
\includegraphics[width=\linewidth]{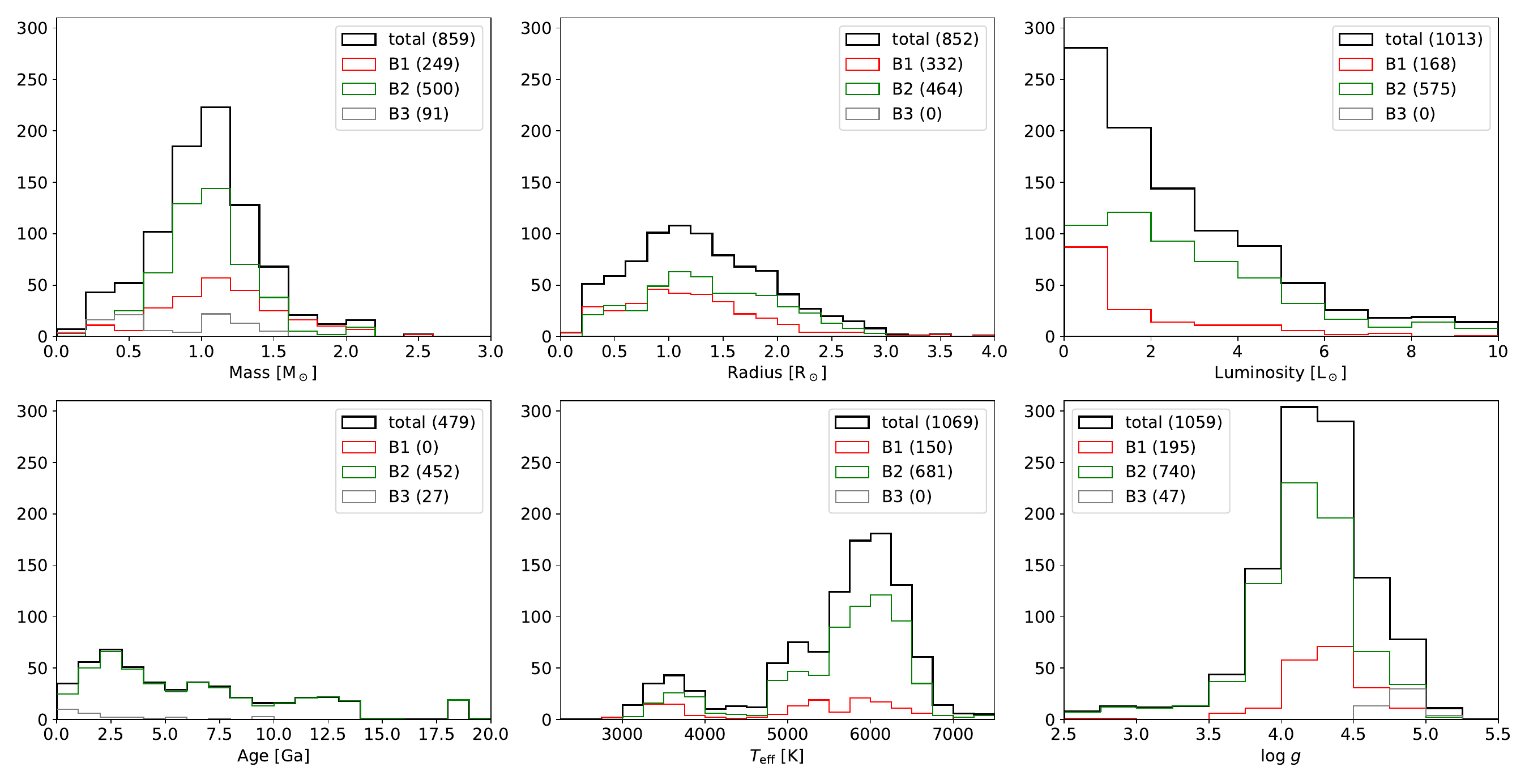}
\caption{Histograms of stellar properties ($M$, $R$, $L$, $\tau$, \teff, \logg) color-coded by benchmark level (B1: red, B2: green, B3: gray). The overall black histogram displays the total distribution of compiled values, including stars without an assigned benchmark tier.}
\label{app:histo}
\end{figure*}

\label{LastPage}
\end{document}